\documentclass[aps,prl,twocolumn,nofootinbib,notitlepage]{revtex4-2}

\usepackage{graphicx,color,tcolorbox}
\usepackage{amsmath,amsthm,amssymb,amsfonts,mathtools,latexsym,cancel,braket}
\usepackage[none]{hyphenat}

\usepackage[pdftex,colorlinks=true,linkcolor=Cerulean,citecolor=Cerulean,urlcolor=Cerulean]{hyperref}
\hypersetup{linkcolor=blue,citecolor=blue,urlcolor=blue}

\usepackage{array,booktabs,longtable,tabularx}
\usepackage{enumitem}
\usepackage{pifont}
\usepackage{changepage}
\usepackage{stackengine}
\stackMath

\DeclareFontFamily{OMS}{oasy}{\skewchar\font48}
\DeclareFontShape{OMS}{oasy}{m}{n}{
	<-5.5>		oasy5     <5.5-6.5>	oasy6
	<6.5-7.5>	oasy7     <7.5-8.5>	oasy8
	<8.5-9.5>	oasy9     <9.5->	oasy10
}{}
\DeclareFontShape{OMS}{oasy}{b}{n}{
	<-6>	oabsy5
	<6-8>	oabsy7
	<8->	oabsy10
}{}
\DeclareSymbolFont{oasy}{OMS}{oasy}{m}{n}
\SetSymbolFont{oasy}{bold}{OMS}{oasy}{b}{n}
\DeclareMathSymbol{\smallleftrightarrow}{\mathrel}{oasy}{"24}

\hypersetup
{
	pdfauthor={
		J. Enrique V\'azquez-Lozano (enrique.vazquez@unavarra.es) \&
		I\~nigo Liberal (inigo.liberal@unavarra.es)},
	pdfsubject={
		Department of Electrical, Electronic and Communications Engineering, Institute of Smart Cities (ISC), Universidad P\'ublica de Navarra (UPNA)},
	pdftitle={A Review on the Scientific and Technological Breakthroughs in Thermal Emission Engineering},
	pdfkeywords={thermal emission engineering, thermodynamics, nanophotonics, quantum optics, thermal radiation, metamaterials, time-varying media}
}

\begin{document}
\title{A Review on the Scientific and Technological Breakthroughs in\\Thermal Emission Engineering}
\author{J. Enrique V\'azquez-Lozano}
\affiliation{Department of Electrical, Electronic and Communications Engineering, Institute of Smart Cities (ISC), Universidad P\'ublica de Navarra (UPNA), 31006 Pamplona, Spain}
\author{I\~nigo Liberal}
\affiliation{Department of Electrical, Electronic and Communications Engineering, Institute of Smart Cities (ISC), Universidad P\'ublica de Navarra (UPNA), 31006 Pamplona, Spain}

\date{\today}

\begin{abstract}
The emission of thermal radiation is a physical process of fundamental and technological interest. From different approaches, thermal radiation can be regarded either as one of the basic mechanisms of heat transfer, as a fundamental quantum phenomenon of photon production, or as the propagation of electromagnetic waves. However, unlike light emanating from conventional photonic sources, such as lasers or antennas, thermal radiation is characterized for being broadband, omnidirectional, and unpolarized. Due to these features, ultimately tied to its inherently incoherent nature, taming the thermal radiation constitutes a challenging issue. Latest advances in the field of nanophotonics have led to a whole set of artificial platforms, ranging from spatially structured materials, and much more recently, time-modulated media, offering promising avenues for enhancing the control and manipulation of electromagnetic waves, from far to near-field regimes. Given~the ongoing parallelism between the fields of nanophotonics and thermal emission, these recent developments have been harnessed to deal with radiative thermal processes, thereby conforming the current basis of thermal emission engineering. In this review, we survey some of the main breakthroughs carried out in this burgeoning research field, from fundamental aspects, theoretical limits, the emergence of new phenomena, practical applications, challenges, and future prospects.
\end{abstract}
\maketitle
\sloppy

\section{Introduction}

Our everyday experience shows that solid bodies, when heated to sufficiently high temperature, become incandescent, namely, they emit radiation within the visible frequency range. This phenomenon is clearly exemplified, for instance, in a furnace, where, as an iron rod is progressively heated, it takes a color which transits from dark red to light yellow, and, at extremely high temperatures, even reaching a bluish-white hue. Importantly, this does not means at all that bodies do not emit radiation at ordinary temperatures, i.e., around~$300$~K. Indeed, all the matter with a temperature greater than the absolute zero emits thermal radiation. However, at room temperature, most of the radiation is emitted in a frequency (wavelength) window lower (higher) than the infrared (IR) range, thus becoming invisible to the unaided human eye. In fact, even at elevated temperatures, e.g., within the order of thousands of kelvin, such as in the stars or in incandescent light bulbs, most of the radiation remains imperceptible to the human eye. Therefore, inasmuch as it is only attributed to the existence of (a difference of) temperature, the emission of thermal radiation is a fundamental and universal physical phenomenon.

An insightful and judicious glance at the three words involved in the own definition of the process of emission of thermal radiation suggests in itself that it actually is an intricate and multifaceted phenomenon, so that it can, and indeed it should, be undertaken from three different perspectives.

Firstly, as it is a {\em thermal} process, thermal radiation can naturally be addressed from the point of view of thermodynamics~\cite{Kittel,Greiner}. From this approach, thermal emission is mostly associated with phenomenological and macroscopic features, as well as with applications related with the optimization of the efficiency of thermal radiative processes such as the generation, transfer, conversion, storage, and retrieval of thermal energy~\cite{Datas}. In this regard, currently there is an upsurging interest in the pursuit of sustainable and efficient techniques for harnessing, managing, and effectively exploiting radiative heat, with a focus on endeavors such as thermal radiation harvesting or waste heat recycling.

Secondly, the term {\em radiation} underscores that thermal radiation is nothing but a propagating electromagnetic wave, and hence, approachable from the point of view of electrodynamics~\cite{Jackson,Landau,Schwinger}. In turn, this suggests the possibility for tackling it under the framework of electromagnetic optics~\cite{Hecht} and nanophotonics~\cite{Novotny}. This standpoint, besides rendering a whole set of platforms and approaches in order to enhance the control over far-field thermal emission features (including the spectral bandwidth, the directivity, or the polarization), have also granted access to a significant number of extraordinary properties and novel effects brought about in the near-field regime. These developments, so far headed by the metamaterials~\cite{Heber2010,Soukoulis2011,Zheludev2010,Zheludev2015}, have played a pivotal role in advancing the integration of thermal emission and photonic engineering upon a common umbrella, thus ushering in unparalleled opportunities to challenge and even overcome some fundamental physical limits. Still, in this overall context, the latest qualitative leap has come with the much more recent proposal of temporal metamaterials (oftentimes simply referred~to as time-varying, or time-modulated, media)~\cite{Engheta2021,Galiffi2022,Yin2022,Yuan2022,Engheta2023}, putting forward a change of paradigm, passing from spatially nanostructuring the geometrical features of matter, to temporally modulating the constitutive properties of the medium. At present, this realization conform the most cutting-edge border in the blooming fields of material science and nanophotonics, and consequently, in thermal emission.

Lastly, the notion of {\em emission}, somehow associated with the process of photons generation, emphasizes the quantum nature of thermal radiation. Indeed, such a quantum character of thermal radiation is straightforwardly evinced from the fact that it was precisely the striving toward a rigorous model to explain this radiative process, specifically, the emission spectrum of the blackbody, what primarily spurred the onset of the quantum theory~\cite{Scully,Loudon,Vogel}. Furthermore, it is exclusively within this quantum formalism wherein the energy contribution associated to the vacuum, or zero-point, fluctuations can be rigorously accounted for and distinguished from thermal fluctuations~\cite{Milonni}. Such a consideration lies on the proper assignment of quantum operators in the correlation functions and their corresponding identification with the processes of annihilation and creation of photons. These realizations for properly addressing the quantum correlations are ultimately rooted within the quantum theory of optical coherence~\cite{Glauber1963,Mandel1965}, which constitutes one of the major milestones of quantum optics~\cite{Mandel}.

The three approaches sketched out above conform the three pillars on which the current state-of-the-art of thermal emission engineering is settled down. Notably, the parallel development with the rapidly evolving field of nanophotonic engineering has boosted a prolific emergence of novel predictions, innovative applications, and hitherto unexplored phenomena in the realm of thermal emission. Upon this basis, here we review some of the main scientific and technological breakthroughs in the field of thermal emission engineering. Specifically, elaborating a little more in-depth discussion on the aforementioned approaches, we pinpoint the forefront fundamental aspects of thermal emission engineering. We~then briefly sketch out two contemporary theoretical frameworks that have paved the way for overcome some constraints. Finally, we present a general outlook on recent technological breakthroughs carried out both from spatial and temporal approaches, discussing the current landscape, practical implementations, and challenges.

\section{Emission of thermal radiation from three different approaches}

\subsection{Thermodynamics approach: Thermal radiation as a source of energy}

\begin{figure*}[t!]
	\centering
	\includegraphics[width=1\linewidth]{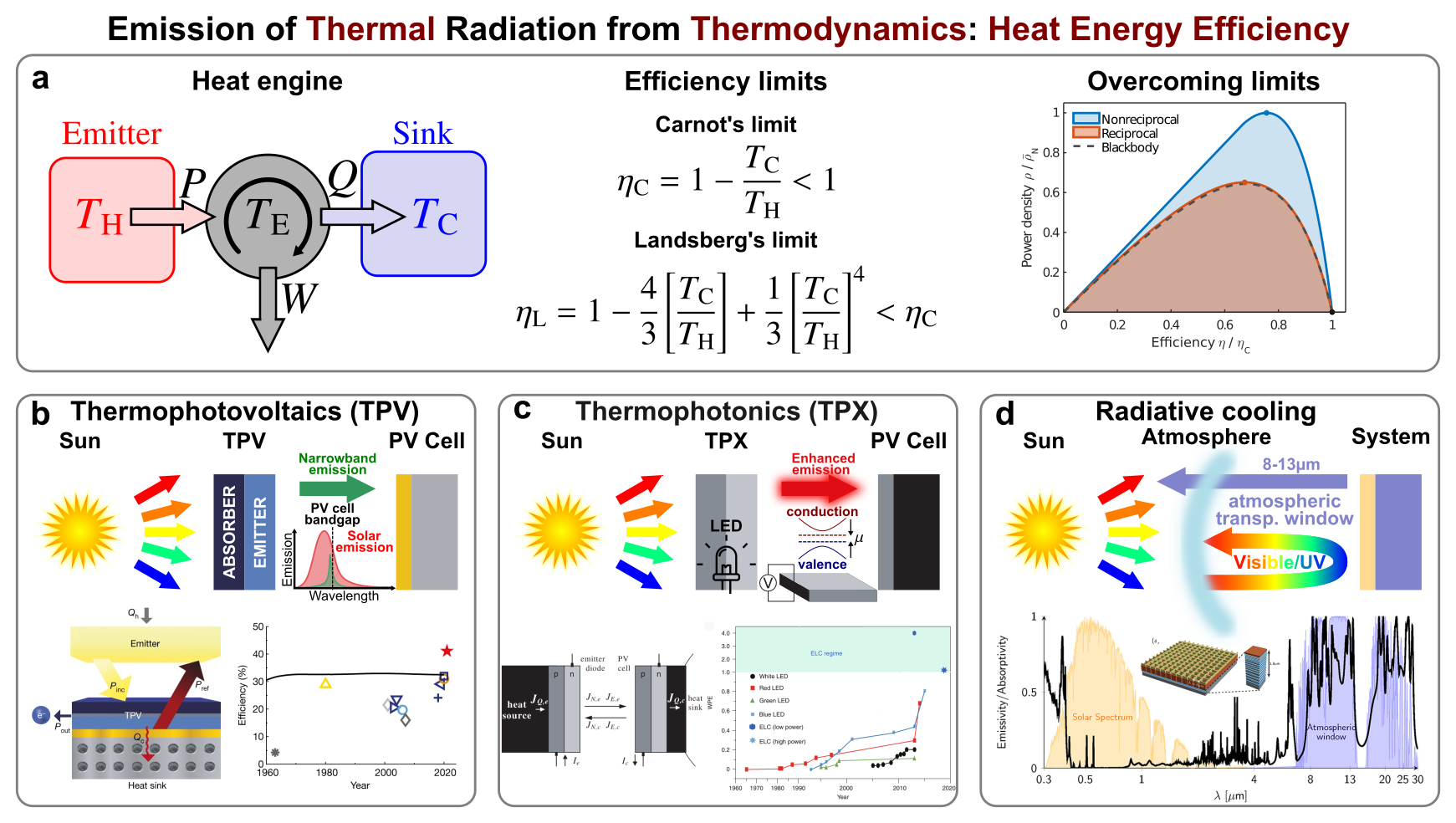}
	\caption{\textbf{Thermodynamic perspective of thermal emission engineering.} (\textbf{a})~Establishing fundamental efficiency limits for the maximum energy harnessing yielded by radiative heat exchange is one of the primary goals of thermodynamics. This involves the design and optimization of systems with the aim of approaching, and even surpassing, theoretical bounds~\cite{Giteau2023A}. Upon this basis, recent research have led to technological advancement such as the development of (\textbf{b})~thermophotovoltaics~(TPV)~\cite{LaPotin2022} and (\textbf{c})~thermophotonics~(TPX)~\cite{Harder2003B,Sadi2020}, as systems to improve the performance of photovoltaic systems. Likewise, (\textbf{d})~radiative cooling has been proposed as a sustainable and efficient technique for heat management~\cite{Rephaeli2013}.}
	\label{Fig.01}
\end{figure*}

From the standpoint of thermodynamics, thermal radiation is introduced, alongside thermal conduction and convection, as one of the basic mechanisms of heat transfer~\cite{Kaviany,Modest,Howell}. Heat, representing a transient exchange of energy between two macroscopic systems at different temperatures, is commonly regarded as a form of energy dissipation, and hence inherently linked to irreversible processes characterized by the generation of entropy~\cite{Tolman1948}. Understanding the entropy production of thermodynamic systems allows to characterize~the~efficiency in energy-heat conversion processes~\cite{Landsberg1980}, and~therefore to establish fundamental upper limits~\cite{Shockley1961,Landsberg1968,Ruppel1980,DeVos1981,Strandberg2015,Buddhiraju2018,Li2020A,Giteau2023A}. Strikingly, this classical theoretical framework still continues to wield a significant influence on the development of current technological applications~[\hyperref[Fig.01]{Fig.~1(a)}].

In particular, in the area of photovoltaics~(PV)~\cite{Green,Luque,Mertens}, there is a well-known theoretical limit, commonly referred to as the {\em Shockley-Queisser limit}~\cite{Shockley1961}, capping the maximum attainable radiative efficiency of solar cell based on single p-n junctions, approximately up to the $30\%$. This limit arises as a consequence of the mismatch between the broadband solar radiation and the band structure of the semiconductors used in solar cells. Noteworthily, the research efforts for overcoming this limit has led to the proposal and the further deployment~of the {\em solar thermophotovoltaic} (TPV) technology~\cite{Swanson1979,Chubb}~[\hyperref[Fig.01]{Fig.~1(b)}]. The working principle of this system for thermal energy harvesting essentially consists in introducing an additional material immediately before the standard PV cell~\cite{Rephaeli2009,Fan2014A,Zhou2016,Wang2022A}. This intermediate element is then to be appropriately engineered to maximize the absorption of radiation across a broad spectral band of solar radiation~\cite{Bierman2016}, thus raising up its temperature by heating, and subsequently emitting the resulting thermal radiation in a narrow~band~right above the bandgap of the solar cell~\cite{Krogstrup2013,Yu2014,Boriskina2014,Boriskina2015,Alharbi2015,Seyf2016,Bermel2010A}. Under ideal conditions, this scheme allows for theoretical efficiencies of up to $85\%$~\cite{Harder2003A}, thereby largely overcoming the aforementioned Shockley-Queisser limit, and even achieving an overall performance near the thermodynamic limit~\cite{Omair2019,Fan2020,Burger2020,LaPotin2022,Lee2022}. At any rate, the fundamental and absolutely unbridgeable limitation in the efficiency of a thermodynamic system~is~ the {\em Carnot's limit}~\cite{Curzon1975,Buddhiraju2018}, being expressed as $\eta_{\rm C}=1-T_{\rm C}/T_{\rm H}<1$, and according to which the maximum efficiency of an ideal heat engine only depends on the ratio between the temperature of the cold ($T_{\rm C}$) and the hot ($T_{\rm H}$) reservoirs. Nonetheless, it should be noted that this is so only for engines where heat exchange~processes are carried out via thermal conduction~\cite{Giteau2023A,Buddhiraju2018}. In the case of radiative heat engines, i.e., when the heat is also exchanged in the form of thermal radiation, the ultimate efficiency is known~as~the {\em Landsberg's limit}~\cite{Landsberg1968,Landsberg1980}, and reads as $\eta_{\rm L}=1-(4/3)(T_{\rm C}/T_{\rm H})+(1/3)(T_{\rm C}/T_{\rm H})^4<\eta_{\rm C}$. By assuming the differential temperature between the Sun ($T_{\rm H}\approx6000$K) and the Earth ($T_{\rm C}\approx300$K), even though the maximum efficiency of an ideal Carnot engine is $\eta_{\rm C}\approx 95\%$, the Landsberg's limit for solar energy conversion just reaches $\eta_{\rm L}\approx 93.3\%$. Importantly, this latter limit can only be approached in nonreciprocal systems~\cite{Buddhiraju2018,Ries1983,Benenti2011,Green2012,Shiraishi2016,Park2021,Park2022A,Park2022B,Giteau2023A}~[\hyperref[Fig.01]{Fig.~1(a)}], i.e., those where the detailed balance of emission and absorption is broken~\cite{Zhu2014A}, e.g., by means of magneto-optic effects~\cite{Ekeroth2017,Zhao2019A}, temporal modulations~\cite{VazquezLozano2023A,Torrent2018}, or in general, by exploiting nonlinearities~\cite{Khandekar2015A}. Here it is worth to stress out that non-reciprocity implies the violation of the Kirchhoff's law of thermal radiation~\cite{Kirchhoff1860,Snyder1998,Greffet1998,Hadad2016,Miller2017,Greffet2018}, but not at all the second principle of thermodynamics, which is actually tied to the time reversibility~\cite{Onsager1931,Miller1960}.

Other than TPV systems, the pursuit of efficient and sustainable techniques to harness and manage solar~\cite{Byrnes2014}, or in general, thermal radiation from any source, aiming to reach or even surpass the thermodynamic performance limits, has led the proposal and the development of other innovative approaches. Two notable and~contemporary examples are {\em thermophotonics} (TPX)~\cite{Harder2003B,Xue2015,Farrel2015} and {\em radiative cooling}~\cite{Hossain2016,Zhao2019B,Zhao2019C,Munday2019,Yin2020,Li2020B,Ahmed2021,Fan2022}.

TPX represents an extension of TPV aimed at improving the thermal-to-electrical energy conversion efficiency by incorporating a light-emitting diode (LED) (or other types of emitters working as a photonic heat engine)~\cite{Harder2003B}, effectively transitioning from a passive thermo-electric to an active electro-luminescence approach for waste heat recovery~[\hyperref[Fig.01]{Fig.~1(c)}]. In contrast to the passive intermediate material in TPVs, the electrically biased LED yields an internal chemical potential, leading to increased radiated power density to the cold side. This enhanced emission can be finely tuned to match the bandgap of the solar cell, enabling TPX systems to operate at significantly lower temperatures compared to TPV~\cite{Oksanen2015}. Likewise, the control of the chemical potential let TPX systems to reverse the heat flow~\cite{Zhao2020,Sadi2020}, thereby allowing them to work as solid-state~refrigerators~\cite{Chen2015A,Chen2016A,Chen2017}.

\begin{figure*}[t!]
	\centering
	\includegraphics[width=1\linewidth]{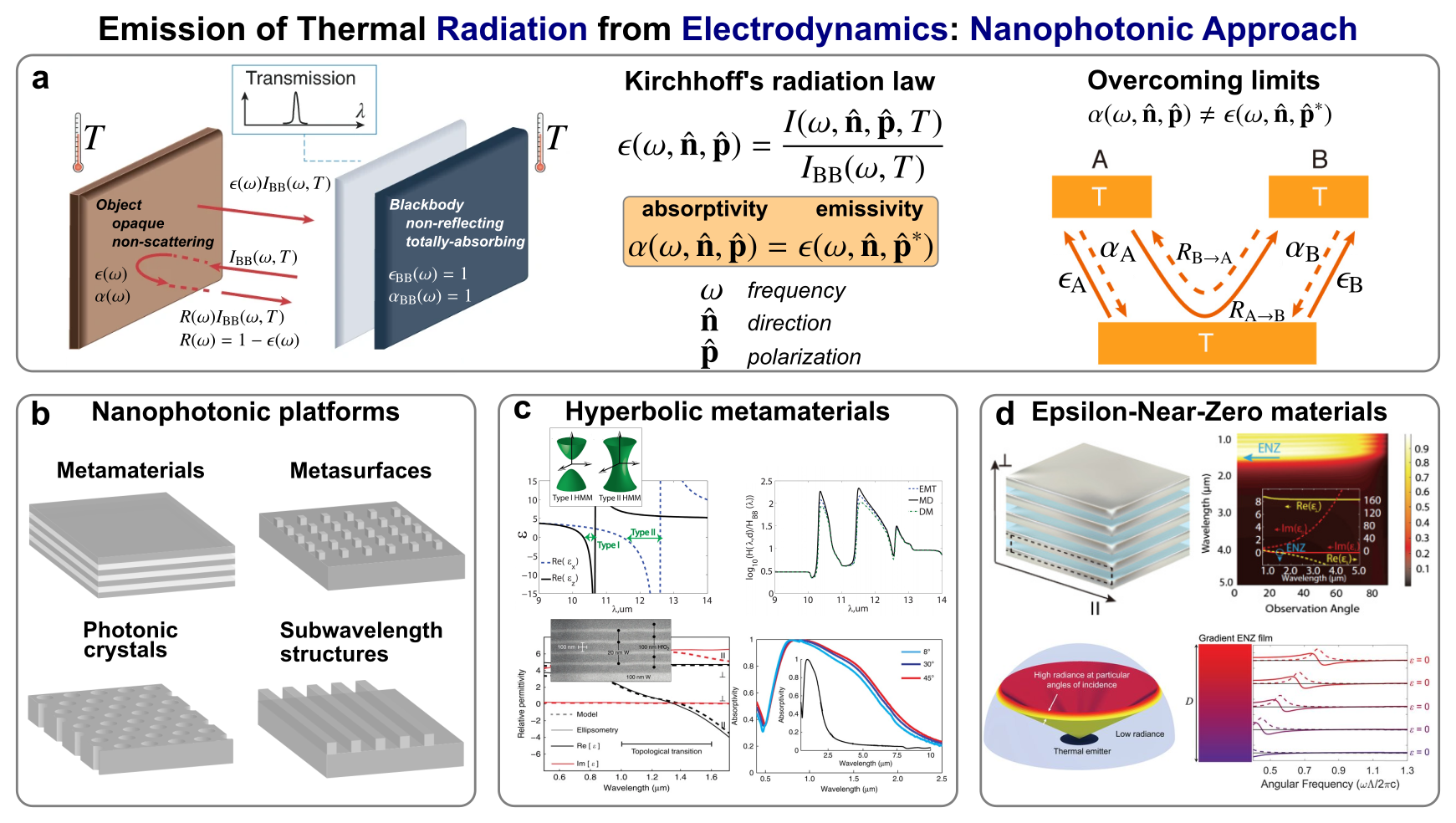}
	\caption{\textbf{Electrodynamic perspective of thermal emission engineering.} (\textbf{a})~Nanophotonic approaches to engineer far-field thermal emission are ultimately underpinned by the Kirchhoff's radiation law, setting an equivalence between emissivity and absorptivity in reciprocal systems~\cite{Li2018Review,Baranov2019Review}. (\textbf{b})~The practical implementation of this theoretical approach has mostly been carried out by means of photonic nanostructures, including metamaterials, metasurfaces, photonic crystals, or subwavelength structures~\cite{Li2021Review}. Upon this ground, regarding their exceptional optical properties, as well as the versatility in the control of material dispersion features, (\textbf{c})~hyperbolic~metamaterials~\cite{Guo2013,Dyachenko2016} and (\textbf{d})~Epsilon-Near-Zero (ENZ) materials~\cite{Molesky2013,Xu2021}, have enabled some of the most relevant advances in thermal emission engineering.}
	\label{Fig.02}
\end{figure*}

On the other side, radiative cooling is a passive technique~\cite{Granqvist1981,Raman2014}, whose working principle is, to some extent, inverse to that of solar TPV devices~[\hyperref[Fig.01]{Fig.~1(d)}]. Regarding the outer space as the ultimate thermodynamic sink (both in terms of extension and temperature, around $3$ K~\cite{Fixsen2009}) where expelling out all the thermal radiation, and noticing that at the typical ambient temperature of the Earth, i.e., around $300$ K, the spectrum of thermal radiation of bodies peaks within the transparency window of the atmosphere, i.e., in the wavelength range of $8-13~\mu{\rm m}$, the idea simply consists in leveraging such a coincidence to resourcefully enhance the efficiency of cooling processes, such as it happens at night~\cite{Chen2016B,Chen2019,Zhu2023}. Upon this basis, the efficiency is optimized by maximizing both the emissivity in the atmospheric transparency window~\cite{Hossain2015}, and, the reflectivity in the entire solar spectrum~\cite{Rephaeli2013}. It should be noted that this latter requirement is equivalent to minimizing the absorptivity (and hence the emissivity) of the solar spectrum, and this is just the opposite condition for maximizing the overall performance of TPV systems, for which solar reflectivity should be minimal~\cite{Kou2017}. Related to this, there are also recent proposals to improve the performance and reliability of solar cells based on radiative cooling as a mechanism to reduce the operating temperature~\cite{Zhu2014B,Safi2015,Zhu2015,Li2017}.

\subsection{Electrodynamics approach: Thermal radiation as an electromagnetic wave}

From a completely different point of view, as a manifestation of a radiative process driven by the propagation of electromagnetic waves, thermal radiation may also be addressed within the framework of electrodynamics (notice that, for now, we are avoiding the specification on classical or quantum)~\cite{Levin}. This connection between thermal properties and~electromagnetic waves is nowadays a commonplace~\cite{Boriskina2016,Boriskina2017}. However, it was not until the beginning of the 19th century when W. Hershel, by performing astronomical spectrophotometry for recording the spectral distribution of stars, discovered the existence of IR radiation and its capability to convey thermal energy~\cite{Herschel1800}. In the course of his research, measuring the temperature of different spectral components of sunlight rays scattered through a prism, Herschel observed an increase in the temperature of the radiation beyond the red part of the spectrum, thereby unequivocally demonstrating the ultimate electromagnetic nature of thermal radiation as a particular mechanism of heat transfer. In such a case, as in the majority of the sources of thermal radiation, and particularly for bodies near room temperature, the emission is given off at the IR frequency range. Notwithstanding, it should be noted that it actually occurs in the entire electromagnetic spectrum. In fact, at very high temperature thermal radiation falls far above the IR range, extending into the visible and even the ultraviolet ranges. Specifically, in the visible range, besides heating, thermal radiation also produces lighting, a phenomenon dubbed as incandescence~\cite{Greffet2011,Polimeridis2015,Ilic2016}.

Be that as it may, inasmuch as thermal radiation behaves as, and indeed it is, an electromagnetic wave, it can propagate indefinitely through a vacuum, in fact reaching there its maximum emission efficiency. Therefore, although with nuances, the main features of the emission of thermal radiation are ultimately encapsulated in the formalism of Maxwell's equations. Under this framework, thermal radiation can be simply thought of as an energy conversion process, where the kinetic (or thermal) energy, due to the continuous and fleeting motion of the atoms and molecules conforming the matter, is transformed into the energy of the electromagnetic fields releasing the body. Here it is important to note that, at finite temperature, any object, even charge-neutral, always owns thermally fluctuating charges, and hence currents (i.e., moving charges), continuously radiating electromagnetic fields. Thus, since brought about by randomly distributed fluctuating electric currents, thermal emission is a purely stochastic process. This explains its inherently incoherent nature, both in space and time, as well as in the degree of polarization. And consequently, this is in turn the reason why thermal radiation generally displays a broadband spectrum, an almost isotropic propagation, and an unpolarized field distribution. In this sense, controlling and enhancing these degrees of coherence constitutes one of the foremost goals of thermal emission engineering.

The electromagnetic nature of thermal radiation has made nanophotonics an exceptionally useful tool for addressing many challenges inherent to thermal emission engineering~\cite{Fan2017Review,Li2018Review,Baranov2019Review,Cuevas2018Review,Li2021Review,Pascale2023A,Chapuis2023,Picardi2023Review}. This encompasses from theoretical frameworks, to practical approaches, including the most cutting-edge set of platforms.

From a theoretical point of view, the nanophotonic approach of thermal emission engineering strongly relies on the control and manipulation of optical material properties~\cite{Ilinskii}. Specifically, the main quantities characterizing thermal emission in the far-field regime are the spectral absorptivity, $\alpha(\omega,{\bf n},{\bf p})$, and the spectral emissivity, $\epsilon(\omega,{\bf n},{\bf p})$. The former accounts for the material absorption of the incoming radiation at a given frequency ($\omega$), direction (${\bf n}$), and state of polarization (${\bf p})$, and is defined as the ratio between the incident and the absorbed electromagnetic power per unit of area. Similarly, the emissivity is characterized as the ratio between the electromagnetic power emanating from a given material at a certain frequency, direction, and polarization, $I(\omega,{\bf n},{\bf p},T)$, normalized with respect to the emission power given off by an ideal blackbody emitter, $I_{\rm BB}(\omega,T)$, both at the same temperature $T$. Noteworthily, in reciprocal optical systems, i.e., those wherein all the involved materials are characterized by means of linear and time-independent symmetric permittivity and permeability tensors (notice that scalar-like isotropic media are the simplest particular case), these two quantities, i.e., absorptivity and emissivity, turn out to be equal. Such an equivalence, commonly referred to as the Kirchhoff's radiation law~\cite{Kirchhoff1860,Snyder1998,Greffet1998}~[\hyperref[Fig.02]{Fig.~2(a)}], is generally expressed as:
\begin{equation}
\alpha(\omega,{\bf n},{\bf p})=\epsilon(\omega,{\bf n},{\bf p}^*)=\frac{I(\omega,{\bf n},{\bf p}^*,T)}{I_{\rm BB}(\omega,T)}\leq1,
\label{Eq.01}
\end{equation}
where the asterisk stands for the complex cojugate polarization, required for the time-reversal operation. According to this statement, the control of the emissivity can be carried out through the absorptivity, and viceversa. This eases the experimental material characterization in terms of these thermal features, which can be directly carried out by measuring the absorptivity. Notice that, direct and accurate measurements of emissivity would require for a precise acquisition of thermal radiation over a sufficiently wide solid angle, and this should be performed by heating up the sample to a temperature high enough so that the signal-to-noise ratio could be detectable, and all in an environment exhibiting transparency at the mid-IR frequency range. Even though it has been demonstrated accessible experimental capabilities to undertake these direct measurements of emissivity~\cite{Dobusch2017,Xiao2019A}, these experiments are quite challenging, and even more in the near-field regime. Thus, indirect measurement of the absorptivity through the reflectivity, $\alpha=1-R$, relying upon a simple energy balance between the emission and absorption, reflection, and transmission (simplified by considering an opaque object, i.e., for which transmission is null), enables a straightforward procedure for determining the emissivity.

Besides the aforementioned practical considerations, the relation provided by Kirchhoff's radiation law also entails fundamental implications. In particular, emissivity and absorptivity are bounded between the limits $0$ and $1$; the former, $\epsilon=0$, characterizing a perfect reflector, and the latter $\epsilon=1$, a perfect emitter, which, by definition, is only theoretically reached in the case of an ideal blackbody, thereby establishing an upper limit on the emissivity independently on the specific frequency, incidence angle, or polarization. Hence, the efficiency of a material as a thermal emitter will determine its efficiency as an absorber of thermal radiation. Likewise, the equivalence displayed in Eq.~\eqref{Eq.01} also enables the computation of the far-field thermal emission spectra of any realistic object, which is simply determined by the blackbody spectrum weighted over the emissivity of the material. In as far as most of the thermal emitters are made of reciprocal materials, this simple relationship constitutes one of the cornerstones underpinning thermal emission engineering under a nanophotonic approach~\cite{Baranov2019Review}.

Upon this theoretical basis, photonic nanostructures have proven to be the most suitable platform for controlling and manipulating the optical properties and light-matter interactions. Over the past few decades, this approach has evolved to encompass thermal emission engineering, boosting a plethora of groundbreaking developments, including novel thermal effects~\cite{Zhou2015,Zhu2016,BenAbdallah2016,Zhu2019,Guo2020A}, and innovative applications~\cite{Schuller2009,Yeng2012,Lenert2014,Catrysse2016,Lee2023}. Indeed, spatially nanostructured photonic materials, characterized by geometric features with sizes at or below the wavelength scale, have garnered significant relevance in the area of thermal emission due to their ability to enhance~far-field thermal emission performance~\cite{Yu2013,Simovski2015}, and~granting access to near-field thermal properties~\cite{Basu2009A,Liu2015A}. Thus far, practical~implementations have mostly~relied on metamaterials~\cite{Joulain2010,Liu2011,Mason2011,Wadsworth2011,Cao2013,Qu2017,Kats2013,Wang2014,Zheng2022,Sklan2018,Xu2019A,Liu2013C,Liu2016,Liu2017A,Biehs2012,Biehs2013,Biehs2015,Tschikin2015,Ding2015,Guo2012,Guo2013,Liu2013A,Liu2013B,Simovski2013,Shi2015,Molesky2013,Jun2014,Dyachenko2016,Campione2016,Niu2018,Xu2021}, metasurfaces~\cite{Argyropoulos2013,Woolf2014,Liu2015B,Costantini2015,Dai2016A,FernandezHurtado2017,Inampudi2018,Li2018A,Cao2018,Wojszvzyk2021,Overvig2021}, photonic crystals~\cite{Lin2000,Pralle2002,Luo2004,Narayanaswamy2004,Lee2005,Laroche2006,Chan2006,Lee2007,Ghebrebrhan2011,Inoue2013,Arpin2013,Kang2019,Rodriguez2011A,Inoue2016,Zoysa2012}, or subwavelength structures~\cite{Inoue2015A}, such as spatial gratings~\cite{LeGall1997,Greffet2002,Laroche2005,Dahan2005,Dahan2007,Chalabi2016,Dai2016B,Messina2017,Guo2016}, cavities~\cite{Maruyama2001,Celanovic2005,Miyazaki2008,Ikeda2008,Ridolfo2013,Khandekar2015B}, or, in general, resonant optical systems~\cite{Gentle2010,Aydin2011,Bermel2011,Wu2014,Liu2017B,Xu2019B,Kathmann2020,Hajduk2021}~[\hyperref[Fig.02]{Fig.~2(b)}]. Among the many realizations carried out in these platforms, for their exceptional properties, there are two examples which have attracted a great deal of attention~\cite{Lobet2023}: hyperbolic metamaterials~\cite{Biehs2012,Biehs2013,Biehs2015,Tschikin2015,Ding2015,Guo2012,Guo2013,Liu2013A,Liu2013B,Simovski2013,Shi2015}, and Epsilon-Near-Zero (ENZ) materials~\cite{Molesky2013,Jun2014,Dyachenko2016,Campione2016,Niu2018,Xu2021}.

Hyperbolic metamaterials~[\hyperref[Fig.02]{Fig.~2(c)}] enables the realization of a special class of highly anisotropic media~\cite{Poddubny2013}. Displaying a hyperbolic dispersion relation, i.e., a permittivity (and eventually a permeability) tensor, $\bar{\varepsilon}$, where one of the diagonal elements exhibits an opposite sign with respect to the other two principal components, they represent uniaxial materials capable of supporting both evanescent and propagating modes. According to this definition, these metamaterials are generally classified in two types~\cite{Guo2013}. Type-I hyperbolic metamaterials, characterized by $\varepsilon_{xx}=\varepsilon_{yy}=\varepsilon_{\parallel}>0$ and $\varepsilon_{zz}=\varepsilon_{\perp}<0$, yield a family of two-unconnected isofrequency surfaces, while type-II hyperbolic metamaterials, characterized by $\varepsilon_{xx}=\varepsilon_{yy}=\varepsilon_{\parallel}<0$ and $\varepsilon_{zz}=\varepsilon_{\perp}>0$, leads to a family of simply-connected isofrequency surfaces. This dual behavior provides an insightful pathway to engineer, and also bridge, both near and far-field thermal features, e.g., by means of optical topological transitions~\cite{Dyachenko2016}. Among the many achievements, this platform has demonstrated extraordinary capabilities in producing near-field induced broadband enhancement of the far-field thermal emission spectra, even exceeding the Planck's limit for the blackbody spectrum~\cite{Nefedov2014,Sohr2019}, as well as in enabling highly directive radiative heat sources~\cite{Barbillon2017,Li2021}.

Closely related with hyperbolic metamaterials, ENZ materials~[\hyperref[Fig.02]{Fig.~2(d)}] have proven to exhibit very exotic wave behavior in nanophotonic~\cite{Liberal2016,Liberal2017A,Liberal2017B,Lobet2020}. In particular, regarding thermal emission effects, the effective stretching of the wavelength inside a ENZ material allows for an intrinsic enhancement of the spatial coherence of thermal fields~\cite{Liberal2018}. However, due to the extreme boundary conditions of ENZ media, the fluctuating thermal currents are completely trapped within the material body, thereby naturally inhibiting~the~releasing of thermal radiation~\cite{Liberal2017C}. Since $Z=\sqrt{\mu/\varepsilon}$, with $\mu$ and $\varepsilon$ being, respectively, the permeability and the permittivity of the medium, another direct consequence of ENZ is the enlargement of the medium's impedance as $\varepsilon\to0$. These high values of the impedance are independent of the geometrical features, thus allowing for ultra-thin film thermal emitters, or absorbers, which have been experimentally demonstrated to display narrowband and stable emission lines~\cite{Navajas2023}. An opposite approach led to the introduction of epsilon-near-pole (ENP) metamaterials~\cite{Molesky2013}. The enhanced thermal emission features, namely, narrowband, omnidirectional, and polarization-independent emission, now enabled by a reduction of the impedance mismatch, have made ENP materials to be ideal candidates for TPV systems, with energy conversion efficiencies able even to exceed the Shockley-Queisser limit. Noteworthily, the complementary approach, has also been investigated, demonstrating both theoretically and experimentally, gradient ENZ materials to enable broadband, polarization-dependent, directional control of thermal radiation~\cite{Xu2021}.

\subsection{Quantum approach: Thermal radiation as a fundamental process of photon production}

\begin{figure*}[t!]
	\centering
	\includegraphics[width=1\linewidth]{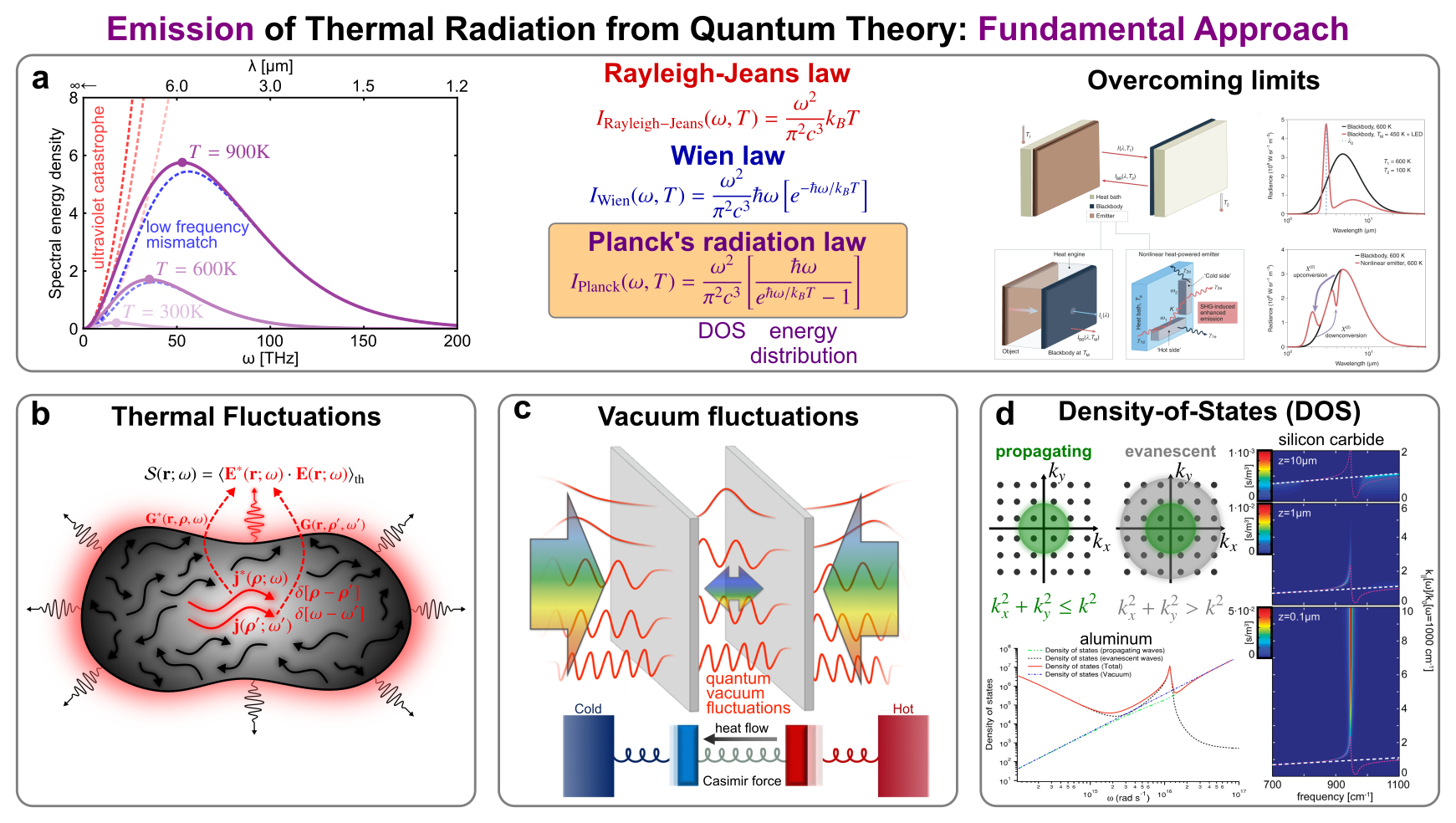}
	\caption{\textbf{Quantum perspective of thermal emission engineering.} (\textbf{a})~After two unsuccessful attempts to classically model the blackbody's thermal emission spectrum, leading to the Rayleigh-Jeans and Wien's laws, the solution came with the realization that, in thermal equilibrium, photon's states are ruled by a quantum statistics, thus yielding the celebrated Planck's radiation law, which sets an upper limit for the far-field radiative heat transfer of macroscopic bodies~\cite{Xiao2022}. (\textbf{b})~The~fluctuational~approach, essentially abridged by the FDT, provides with a more realistic a complete theoretical framework for thermal emission~\cite{VazquezLozano2023A}. (\textbf{c})~Besides thermal fluctuations, the merging of fluctuational and quantum electrodynamics allows one to cope with zero-point quantum vacuum fluctuations~\cite{Gong2021,Fong2019}. (\textbf{d})~Other than fundamental aspects, the quantum perspective of thermal emission offers a practical vision to engineer the density of states~\cite{Joulain2003,Jones2012}.}
	\label{Fig.03}
\end{figure*}

Finally, and as previously anticipated, the emission of thermal radiation is a photon production process, and consequently it is ultimately raised on the quantum theory~\cite{Planck1901,Planck}. Although this fact is nowadays well-understood, and even trivially assumed, at the beginning of the last century, such a realization was not so obvious at all~\cite{Rayleigh1905A,Rayleigh1905B,Jeans1905}. Indeed, first attempts for modeling the thermal emission spectrum of the blackbody were initially based on classical arguments.

The notion of a blackbody, originally put forward in 1860 by G. Kirchhoff~\cite{Kirchhoff1860}, abridges the idea of a non-reflective and totally-absorptive physical object in thermal equilibrium, regardless of the radiation features, i.e., the frequency, direction, and polarization state. This ideal system, regarded as a perfect emitter (or, in virtue of the Kirchhoff's radiation law, a perfect absorber), establishes an upper limit in thermal emission~\cite{Xiao2022}, so that, in general, real materials only emit a fraction of such a blackbody radiation, which is determined by the emissivity (or, equivalently, through the absorptivity)~[cf.~Eq.~\eqref{Eq.01}]. Furthermore, on account of the second law of thermodynamics, this simplified approach allows to unveil a crucial feature: in thermal equilibrium, the shape of the thermal emission spectra only depends on the temperature of the emitter body, independently of the geometry or the material properties. Despite being an idealization, there are many examples of realistic systems comporting as a blackbody, from the typical cavity with a tiny hole, whose blackbody behavior was experimentally demonstrated by O. Lummer and~E.~Pringsheim in 1898~\cite{Lummer1899}, to near-black materials~\cite{Chun2008,Yang2008}, or stars~\cite{Suzuki2018,Serenelli12019}.

Building upon this simple model, and after several qualitative assessments rooted on purely classical arguments, in 1900, it was put forward what we nowadays refer to as the {\em Rayleigh-Jeans law} for the spectral energy density (i.e., the energy per unit volume) of the blackbody radiation:
\begin{equation}
I_{\rm Rayleigh-Jeans}(\omega,T)=\frac{\omega^2}{\pi^2c^3}k_BT,
\label{Eq.02}
\end{equation}
where $k_B$ and $c$ stand, respectively, for the Boltzman constant and the speed of light in vacuum. While being a good approximation either at the low-frequency or the infinite-temperature limits (i.e., at $\omega\to0$ and $T\to\infty$), this statement fails in correctly providing the form of the blackbody spectral distribution~[\hyperref[Fig.03]{Fig.~3(a)}], leading to so misleading predictions such as the divergent emission of high-frequency radiation beyond the ultraviolet range, a behavior that was dubbed by P.~S. Ehrenfest as the \textit{ultraviolet catastrophe}~\cite{Ehrenfest1911}.

Chronologically a bit earlier, in 1896, W.~Wien, based on the theoretical work carried out by L.~Boltzmann and the experimental results obtained by J.~Stefan, originally proposed what he thought of as a complete description for the spectrum of thermal emission:
\begin{equation}
I_{\rm Wien}(\omega,T)=\frac{\omega^2}{\pi^2c^3}\hbar\omega\left[e^{-\hbar\omega/k_BT}\right].
\label{Eq.03}
\end{equation}
Notice that the fundamental constants, in particular,~$\hbar$, being the reduced Planck's constant, were later introduced. This formula is commonly referred to as the {\em Wien's law}, and should not to be confused with the more familiar Wien's displacement law, which states that the shifting of the emission peak with respect to the temperature, is inversely proportional to the wavelength, or directly proportional to the frequency. At any rate and again, as theoretically rooted on a classical framework, this statement fails in fitting with experimental data, in this case, in the low-frequency limit~[\hyperref[Fig.03]{Fig.~3(a)}].

Then, it was not until the groundbreaking proposals and derivations put forward by M. Planck in 1900 that a comprehensive framework for thermal radiation was completely and rigorously established (notice that first derivation provided by Planck was actually empirically obtained by means of a suitable fitting with experimental data), solving at once both discrepancies, at low and high frequency limits~\cite{Jeans1905}. Planck's seminal contribution raised on the fact that the emission of~thermal radiation occurs via discrete energy packets~\cite{Planck1901,Planck}, nowadays referred to as photons (at that time known as ``\textit{quanta~of~light}''), ultimately giving rise to a sound and accurate expression for the blackbody thermal emission~[\hyperref[Fig.03]{Fig.~3(a)}], which is encapsulated within the renowned {\em Planck's radiation law}:
\begin{equation}
I_{\rm Planck}(\omega,T)=\frac{\omega^2}{\pi^2c^3}\hbar\omega\left[\frac{1}{e^{\hbar\omega/k_BT}-1}\right].
\label{Eq.04}
\end{equation}
As much as Kirchhoff's radiation law [cf. Eq.~\eqref{Eq.01}], but even far beyond the matter of thermal radiation, this expression, and more particularly, its derivation, is of paramount importance in the entire field of physics, since it is largely considered as one of the cornerstones that spurred the onset of the quantum theory.

From the above expressions, it can be appreciated~at~glance a similar form: a common prefactor, ${\rm DOS}=\omega^2/(\pi^2c^3)$, accounting for the density of states, followed, in each case, by different functions characterizing the corresponding average energy distribution of the harmonic oscillators as a function of temperature. In this regard, it should be noted that the derivation of the latter expressions strongly relies on the consideration that the radiation can be treated as a photon gas, which is in turn modeled as a harmonic oscillator ensemble. Importantly, what distinguishes the different models from each other, ultimately determining the quantum character of the Planck's law, resides on the characterization of the condition of thermal equilibrium. In the former two classical approaches, namely, those leading to the Rayleigh-Jeans and the Wien's laws, the thermal equilibrium relied, respectively, on the {\em equipartition theorem} and the {\em Maxwell-Boltzmann velocity distribution}, both stemming from the classical kinetic theory. However, these approaches naively extrapolates insights from molecular gases into the behavior of a photon gas, therefore, leading to the aforementioned misleading outcomes. Indeed, important differences such as the constant speed of photons, their non-self-interactive character (at low energies), or the fact that the photon numbers are not conserved (even in a closed system), led Planck to break away from the existing paradigm and make his bold assumption, namely, that thermal equilibrium is characterized by a frequency distribution of harmonic oscillators wherein they can only take up  discrete amounts energies $\hbar\omega$. Based on these arguments (in reality some other more complicated involving thermodynamic and statistical considerations concerning the energy and entropy), Planck debunked the idea that the energy levels form a continuum, and demonstrated that photons actually obey a quantum statistics, the {\em Bose-Einstein~distribution}, from which it is said that photons are bosons. Noteworthily, this new framework encompasses at a time both the Rayleigh-Jeans and Wien's laws, each in the corresponding limit, as well as the Wien's displacement and the Stefan-Boltzmann laws, according to which the total radiated power is given by $P=\sigma T^4$, where $\sigma$ stands for the Stefan's~constant~\cite{Stefan1879}. Likewise, either in the limit $\hbar\to0$ or $\hbar\omega\ll k_BT$, i.e., dropping out the term accounting for the quantum character of thermal radiation, the classical limit determined by the Rayleigh-Jeans law is recovered. Notwithstanding the foregoing, it is also worth pointing out that, since then, there have been other proposals to rederive the blackbody radiation spectrum under classical approaches~\cite{Boyer1969,Boyer2018}.

Other than the foundational basis of thermal radiation, the quantum approach has also paved the way toward the much more recent development of fluctuational electrodynamics~\cite{Levin,Rytov}. Roughly speaking, this approach provides a kind of bridge between both the classical and the quantum formalism for thermal radiation, ultimately determined by the manner in which the radiation is described, whether in terms of classical electromagnetic fields or by means of quantum operators~\cite{Agarwal1975}. More importantly, this fluctuational treatment has ushered in a really insightful pathway to cope with the emission of thermal radiation in a more realistic and complete fashion~\cite{VazquezLozano2023A}. In this regard it is worth emphasizing that thermal equilibrium is an idealization, practically never encountered in real physical systems. Furthermore, the own dynamical character of the dissipation process of thermal emission is, a priori, intrinsically incompatible with the existence of thermal equilibrium. In this sense, whereas Planck's radiation law is only strictly valid for systems at thermal equilibrium, providing with a reasonable approximation in the far-field regime, the fluctuational approach allows for extending the treatment to the near-field regime~\cite{Francoeur2008A}, at the same time that relaxes the underlying global equilibrium condition, only requiring for it to be local. Upon this assumption, and noticing that in thermal equilibrium, even though the system's properties are in a steady state they can still fluctuate around their mean values, the fluctuational approach comes to put together at once both the notions of a non-equilibrium system displaying a dynamical and dissipative behavior, alongside the condition of thermal equilibrium. And, this is essentially made on the basis of the linear-response theory, whereby it can be shown that the source of the~fluctuations is very closely related with the losses~\cite{Nyquist1928,Callen1951,Callen1952}. This statement is encapsulated within~the~so-called {\em fluctuation-dissipation theorem}~(FDT)~\cite{Kubo1966,Eckhardt1982}, which constitutes one of the fundamental pieces of statistical physics with far-reaching implications. It establishes a general relationship between the rate of the dissipated energy in a non-equilibrium system to the correlations of random and fleeting fluctuations that spontaneously and continuously appears at different times and locations in equilibrium systems~\cite{Kruger2011,Rodriguez2012,Rodriguez2013}. So, in the particular case of a material body at finite temperature, the FDT relates the correlations of the thermally fluctuating electromagnetic currents (i.e., those resulting from the thermally induced random motion of charged particles inside the hot body), described within the framework of the classical electrodynamics (and hence relying on the formalism Maxwell's equations, including the presence of source current densities), with the spectral density of thermal radiation, expressed out through the electromagnetic field correlations~[\hyperref[Fig.03]{Fig.~3(b)}]. Herein, it should be noted that the dissipative features of the system, namely, those actually yielding the process of thermal emission, are encompassed within the macroscopic constitutive relations, i.e., the electric permittivity and magnetic permeability, specifically, in compliance with the Kramers-Kronig relations (which in turn underpin  the fundamental principle of causality), by their dispersive properties~\cite{Landau,Jackson,Schwinger}. Precisely owing to this neat characterization, the fluctuational approach allows for analyzing and completely addressing the emission of thermal radiation and its features both in the far and the near-field regime, thereby extending the Planck's law.

Furthermore, when thermal emission is modeled within the framework of quantum electrodynamics, it allows to deal with zero-point quantum vacuum fluctuations~\cite{Riek2015}. Thus, besides enabling a rigorous and unified treatment for addressing and distinguishing both thermal and quantum vacuum fluctuations~\cite{VazquezLozano2023A}, the approach based on quantum electrodynamics opens the door to a new class of striking quantum effects~\cite{Joulain2005,Bimonte2009A,Messina2011,Kruger2012,Fong2019}. Likely, the most paradigmatic example lies on the Casimir effect~\cite{Casimir1948,Lamoreaux1997,Dalvit,Jaffe2005,Rodriguez2011B,Woods2016,Gong2021}, accounting for the appearance of a long-range vacuum-induced dispersion force between two neutral bodies~\cite{Mahanty,Buhmann2007}, due to the quantum fluctuations of the electromagnetic field~[\hyperref[Fig.03]{Fig.~3(c)}]. Even though these forces cannot be ignored at the nanoscale, their effects are generally very weak at macroscopic scales. Yet, relatively recent progresses in nanophotonic and quantum engineering have showed the potential of different dynamical mechanisms for amplifying quantum vacuum fluctuations, up to levels that even enable the extraction of photons from the vacuum state~\cite{Nation2012}. Main examples of such vacuum amplification effects are the parametric amplification~\cite{Renger2021}, and the celebrated dynamical Casimir effect~\cite{Dodonov2020}. Alongside these effects, the scope of the FDT may concern to other related quantum phenomena, for example, non-contact quantum friction~\cite{Volokitin1999,Kardar1999,Volokitin2011A,Silveirinha2014}, whose existence at the absolute zero-point of temperature has authoritatively been questioned~\cite{Pendry1997,Philbin2009,Volokitin2011B,Philbin2011,Pendry2010A,Leonhardt2010,Pendry2010B}. Far beyond these ontological controversies, the quantum friction, and its relation with thermal fluctuations~\cite{Zurita2004,Volokitin2007,Intravaia2014,Intravaia2016A,Intravaia2016B,Intravaia2019,Reiche2020,Oelschlager2022}, has been extensively investigated in multitude of configurations~\cite{Manjavacas2010A,Manjavacas2010B,Zhao2012,Manjavacas2017,Maghrebi2012,Maghrebi2014}.

Besides these fundamental aspects, from the aforementioned fluctuational perspective, ultimately relied on the FDT, the quantum approach of thermal radiation offers an insightful and practical vision to control and manipulate the thermal emission spectra. Indeed, just like from the electrodynamics approach the spectral properties of thermal emission are determined by the materials' emissivity (or, by virtue of the Kirchhoff's radiation law, the absorptivity), from the quantum standpoint, such a control can be alternatively carried out by means of the density of states~(DOS)~engineering~\cite{Jacob2010,Joulain2003,Joulain2005,Francoeur2010,Messina2013,Jones2012,Jones2013,Babuty2013,OCallahan2014,Chen2015B}. Both approaches are related to each other via the Green's function formalism~\cite{Sipe1987,Francoeur2009,Narayanaswamy2014}. Be that as it may, the underlying idea of DOS engineering lies in the possibility to modifying the number and the distribution of available thermal photonic states, and accordingly, the amplitude and the shape of the resulting thermal emission spectra~[\hyperref[Fig.03]{Fig.~3(d)}]. Notice that in Eqs.~\eqref{Eq.02}-\eqref{Eq.04}, the prefactor ${\rm DOS}=\omega^2/(\pi^2c^3)$ corresponds to the case of free-space thermal photons propagating in the far-field regime. Thus, if instead of vacuum, the emitter is embedded in an arbitrary (regardless of its dispersive and absorptive features) medium, the density of states changes. Likewise, in the near-field, where the dominant contribution is led by evanescent modes~\cite{Novotny}, thermal fields are both sharply localized nearby the emitter and rapidly decaying away from it. So, inasmuch as in this regime the radiation, and hence the density of states, strongly depends on the geometrical features of emitter (including both the distance and the structural size, shape, and orientation)~\cite{Carminati1999,Shchegrov2000,Mulet2010,Edalatpour2016}, it is typically referred to as the local density of states (LDOS)~\cite{Joulain2003,Francoeur2010,Messina2013}. Apart from modifying the spectral distribution, LDOS gives access to additional channels over the frequency-wavevector space, allowing so for a large amplification of the spectral emissivity. In this regard, materials supporting surface resonant modes have proven to play a crucial role. Typical examples are polar dielectrics or metallic structures, supporting, respectively, surface phonon-polaritons (SPhPs), and surface plasmon-polaritons (SPPs)~\cite{Low2017}. The excitation of such surface modes produces a large increase of the DOS~\cite{LeGall1997,Joulain2005,Joulain2003}, and consequently, yield resonant-induced enhanced thermal emission. However, this effect is only manifest in the near-field, and turns drastically faded in the far-field regime~\cite{Pascale2023B}. Still, as has been theoretically and experimentally demonstrated, it is possible to extract such an enhanced contribution from the near-field and couple it to free-space radiation, e.g., with the aid of photonic nanostructures~\cite{Fan2017Review,Li2018Review,Baranov2019Review,Cuevas2018Review,Li2021Review,Pascale2023A,Chapuis2023}.

\section{General aspects of thermal~emission~engineering}

\begin{figure*}[t!]
	\centering
	\includegraphics[width=1\linewidth]{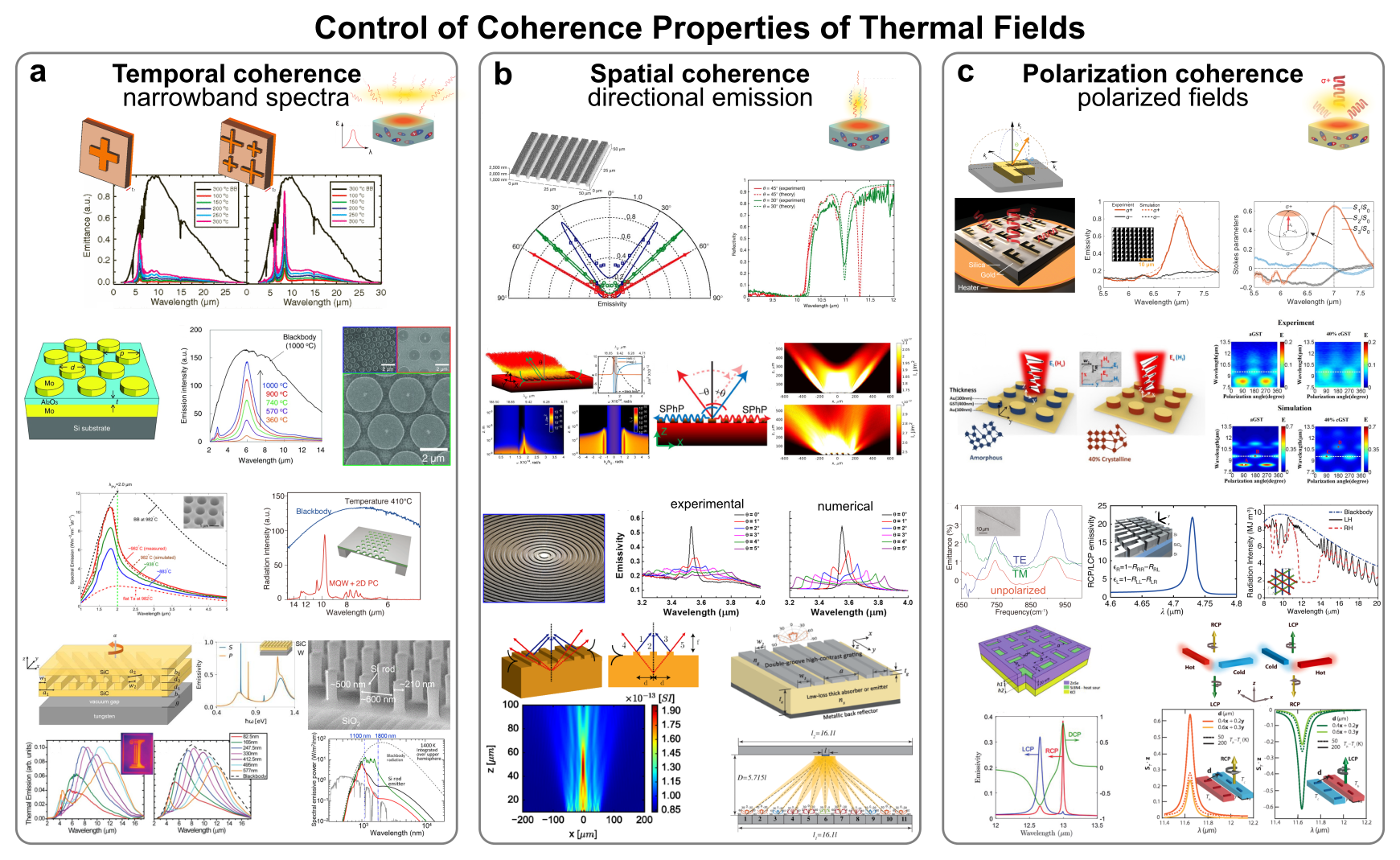}
	\caption{\textbf{Photonic nanostructures to control coherence properties of thermal fields.} Palette of spatially engineered photonic nanostructures used to control and enhance the degree of (\textbf{a})~temporal coherence~\cite{Liu2011,Yokoyama2016,Rinnerbauer2013,Zoysa2012,Guo2021,Streyer2013,Asano2016}, (\textbf{b})~spatial coherence~\cite{Greffet2002,Inampudi2018,Park2016,Chalabi2016,Sakr2017}, and (\textbf{c})~polarization coherence~\cite{Wang2023,Qu2018A,Schuller2009,Wu2014,Lee2007,Dyakov2018,Khandekar2019}, each of them respectively related with the spectral bandwidth, the directivity, and the state of polarization of thermal radiation.}
	\label{Fig.04}
\end{figure*}

So far, we have shallowly outlined how several branches of physics, specifically, thermodynamics, electrodynamics, and the quantum theory, has mutually impacted in the definite conformation of the contemporary area of thermal emission engineering. We have seen that, whereas the thermodynamic approach has greatly motivated the development of thermal emission engineering for technological energy applications, the quantum theory has provided with a fundamental framework setting down the basic theoretical foundations. Thus, within this schematic scenario, and on the basis of the latest advances performed in the field of nanophotonic engineering, the electrodynamic approach can be though of as a kind of bridge between both perspectives, whose central pillar, gathering together both fundamental aspects and practical implementations in realistic platforms at once, lies on the notion of optical coherence.

\subsection{Coherence properties of thermal fields}

Broadly speaking, coherence is one of the most distinctive characteristics of waves~\cite{Glauber1963,Mandel1965,Mandel}, which, regardless of their nature~\cite{Mehta1965}, describes their capability to produce interference. In other words, it sets down a metric to determine the statistical similarity of a wave with respect to a given parameter. This property has widely proven its utility in areas of physics concerned with electromagnetic fields~\cite{Landau,Jackson,Schwinger}, from classical optics and nanophotonics~\cite{Hecht,Novotny}, to quantum optics~\cite{Scully,Loudon,Vogel}. Thus, being a particular form of electromagnetic wave, it can of course be applied to thermal radiation. By conveniently adopted, specifically, by distinguishing between temporal, spatial, and polarization coherence, one can estimate the average correlations between (thermal) fields at different instants of time, spatial positions, or in distinct polarization states. Remarkably, by means of these correlations, the different types of coherence can be correspondingly related to waves features, such as the {\em spectral bandwidth}, the {\em directivity}, or the {\em state of polarization}, respectively.

Within this context, the main difference between thermal radiation and (non-thermal) light, customarily undertaken in optical and nanophotonic systems, relies on the degree of coherence of the source. So, whereas conventional photonic sources, such as lasers or antennas, produce coherent light, the inherently stochastic nature of thermal sources (i.e., the finite temperature of hot bodies) leads to totally uncorrelated (thermally) fluctuating electromagnetic currents, which makes thermal fields to be highly incoherent. Hereupon, thermal radiation is typically characterized to display a broadband spectra, omnidirectional field distributions, and unpolarized propagation. Thus, just like photonic nanostructures have enabled an enhanced control over these features of light emanating from coherent sources, they can also be seized for taming thermal radiation, both in the far~\cite{Fan2017Review,Li2018Review,Baranov2019Review}, and the near-field regimes~\cite{Cuevas2018Review,Li2021Review,Pascale2023A,Chapuis2023}.

\subsubsection{Temporal coherence: Spectral bandwidth}

Through the convolution theorem, relating the Fourier transform of the thermal emission spectrum with the autocorrelation function, the temporal coherence of thermal radiation can be directly linked with their spectral features, specifically, with the bandwidth or the occurrence of sharply localized resonant peaks attributed to the excitation of surface modes. Typical approaches to tune and enhance these spectral features rely on the tailoring of the emissivity, which, from the nanophotonic framework, can be directly performed through material dispersion engineering. In this sense, as reported in the recent literature~\cite{Liu2011,Zoysa2012,Inoue2013,Guo2016,Inoue2016,Liu2017B,Pralle2002,Dahan2007,Aydin2011,Yeng2012,Mason2011,Inoue2015A,Yokoyama2016,Rinnerbauer2013,Guo2021,Streyer2013,Asano2016,Yang2017,Lochbaum2017,Puscasu2008,Liu2015C,Zhao2014,Kohiyama2015,Matsuno2017,Sakurai2019,Inoue2015B}, spatially engineered photonic nanostructures have paved the way toward a wealth of advantageous possibilities for controlling temporal coherence of thermal fields.

In~\hyperref[Fig.04]{Figure~4(a)}, we showcase various representative examples of different photonic nanostructures used to control the spectral response of thermal radiation, namely, both the amplitude and the spectral bandwidth. In particular, we can see how different arrays of resonant structures with a wide variety of geometrical shapes and sizes can appropriately be designed to enable narrowband far-field thermal emission spectra. One of the most enlightening and pioneering examples is that put forward in 2011 by Liu and colleagues~\cite{Liu2011}, and subsequently highlighted by Greffet~\cite{Greffet2011}. There it is proposed a metal-insulator-metal (MIM) structure consisting in a periodic array of cross-shaped metallic resonators separated from a metallic mirror by a transparent, thin, dielectric substrate. Behaving as a metamaterial perfect absorber, this structure allows for the possibility to yield near total absorptivity (and hence emissivity) in a narrow spectral bandwidth. Interestingly, by combining differently designed crosses in a single array, thereby enabling multiple resonances, the structure can produce dual-band spectrally narrow thermal emission. Other proposals either based on~MIM metamaterials~\cite{Yokoyama2016} or in photonic crystals~\cite{Rinnerbauer2013,Zoysa2012} have shown excellent performance characteristics in terms of spectral selectivity and stability. Furthermore, it has also been presented more complex designs enabling narrow-band thermal emission with tunable both over a wide frequency range and at arbitrary temperature~\cite{Guo2021}, strong absorption and selective thermal emission achieved by the combination of engineered mid-IR metals coated with subwavelength high-index dielectric layers~\cite{Streyer2013}, as well as enhanced and selective thermal emission mechanism based on the intertwined action of interband transitions and the Mie resonances enabled by an array of silicon nanorods~\cite{Asano2016}.

As seen above, the comprehension and the possibility to engineer the resonant behavior of structures offers a valuable conceptual pathway for engineering the spectral response of thermal emitters. Besides the conventional approach of enhancing thermal emissivity, another alternative consists in actively tailoring the suppression of the emissivity of a material over specific frequency ranges~\cite{Cornelius1999}. This can be achieved, for instance, by means of bandgap engineering in photonic crystal structures. Noteworthily, there are nanophotonic structures where these two approaches can be simultaneously implemented, exhibiting so the dual capability of amplifying emissivity in certain frequency ranges while being suppressed in others~\cite{Lin2000}. Notwithstanding the foregoing, and as previously pointed out, it should be noted that, regardless the geometrical features of the structure, or the material dispersion, since spatially structuring is a passive approach, namely, not involving the pumping of extra energy into the system, the blackbody emission spectrum constitutes an absolutely and unbridgeable upper limit.

\subsubsection{Spatial coherence: Directivity}

Just like temporal coherence is related with spectral features, the spatial coherence, is related with the directivity, i.e., with the angular selectivity. In this case, typical approaches to control this property are majorly based on the geometrical aspects of material structures. In this regard, as reported in the recent literature~\cite{Greffet2002,Inampudi2018,Chalabi2016,Costantini2015,Dahan2007,Argyropoulos2013,Shen2014,Park2016,Sakr2017,Chung2017,Zhou2021}, artificially shaped photonic nanostructures have again proven to be a suitable platform for controlling the direction of propagation of thermal fields.

In~\hyperref[Fig.04]{Figure~4(b)}, we show some representative structures and configurations to tailor the angular or directional properties of thermal radiation. In this case, doubtless, the most groundbreaking example is that put forward in a seminal work by Greffet and colleagues in 2002~\cite{Greffet2002}, consisting in a subwavelength grating structure made of silicon carbide (SiC), from which they experimentally demonstrate highly directional coherent thermal radiation. Here, the Rayleigh anomaly of a grating enables the resonant excitation and the ensuing diffraction of surface modes (specifically, SPhPs), thereby extracting the enhanced contribution from near-field, to then couple it into far-field radiation propagating in free-space. This extraction-coupling process can be simply described by means of the {\em momentum-matching~condition}:
\begin{equation}
k_0\sin{\theta}=k_\parallel+mG,
\label{Eq.05}
\end{equation}
where $k_0=2\pi/\lambda$ stands for the free-space wavenumber, $\theta$~is the emission angle, $k_\parallel$ is the wavenumber of the surface mode, $m$ is an arbitrary integer number denoting the diffraction order, and $G=2\pi/d$ is the grating reciprocal vector, with $d$ being the grating period. Ideally, this should occur for a continuum of frequencies supporting a given transversal wavenumber fixing the direction of propagation. However, the dispersion of the specific material (e.g., that of SiC), affecting to the temporal coherence, is to be accounted for as well. Hence, angular and frequency selectivity (i.e., spatial and temporal coherence) are often simultaneously achieved~\cite{Costantini2015}. Despite that, it is worth emphasizing that there are several works reporting different strategies to independently control the directivity of thermal~radiation over broad spectral bandwidths~\cite{Argyropoulos2013,Chung2017,Xu2021}. Recently, another proposal based on periodically patterned SiC metasurface, has theoretically investigated the possibility to unidirectionally routing the thermal radiation~\cite{Inampudi2018}. Likewise, more intricate designs, such as bull's eye structures made of tungsten~(W) and molybdenum~(Mo), have been proposed and experimentally demonstrated to enable highly directional thermal emission~\cite{Park2016}. Furthermore, other interesting possibilities consist in exploiting the versatility and flexibility of metasurfaces to manipulate the phase for inducing non-uniform phase gradients. Upon this idea, it has been shown that, by properly engineering the geometrical features of a metasurface, one can produce scattering of thermally excited surface modes, so that they can be out-coupled from the surface, interfere constructively, and yielding free-space focusing, thereby mimicking the behavior of a lens for thermal radiation~\cite{Chalabi2016}. Following a similar principle, there has also been proposed an angle-selective reflective filter to efficiently and selectively reduce, or even suppress, thermal emission in a certain spatial~region~\cite{Sakr2017}.

\subsubsection{Polarization coherence: State of polarization}

Lastly, another coherence property that can be manipulated in thermal fields concerns the state of polarization. In this case, polarization coherence is related with the fact that orthogonally polarized electromagnetic waves do not interfere to each other. Likely due to the more subtle repercussion on practical applications, in comparison with the previously discussed temporal and spatial coherence, little attention has been paid to the control of this property, which is otherwise fundamental for electromagnetic fields. Nonetheless there are still a number of works tackling on different approaches to control the polarization coherence of thermal fields~\cite{Schuller2009,Wu2014,Lee2007,Miyazaki2008,Ikeda2008,Dahan2005,Wadsworth2011,Abbas2011,Qu2018A,Dyakov2018,Khandekar2019,Wang2023,Wojszvzyk2021,Bimonte2009B,Ohman1961,Sandus1965,Liu2022,Wang2022B}. In this regard, the generation, manipulation, and mutual conversion of linear, circular, or arbitrary elliptical polarization, is essentially achieved by means~of~photonic nanostructures that break some symmetries~\cite{Liu2022,Wang2022B,Wang2023}, either geometrical (i.e., involving anisotropic~\cite{Cui2012A}, non-periodical~\cite{Zhang2021}, or chiral structures~\cite{Cui2012B}), modal (i.e., involving non-Hermiticity~\cite{Doiron2019}, or asymmetrical resonances~\cite{PerezRodriguez2017}), or global (i.e., involving nonreciprocal~\cite{Guo2020B}, irreversible~\cite{Ghanekar2022}, or nonlinear~\cite{Soo2016} optical systems).

In~\hyperref[Fig.04]{Figure~4(c)}, we show some illustrative examples of structures to manipulate the polarization properties of thermal radiation. In this regard, we highlight a very recent work carried out by Jacob and colleagues~\cite{Wang2023}, where, by means of symmetry-broken metasurfaces, they experimentally show that spinning (i.e., circularly polarized) thermal radiation with a nonvanishing optical helicity can be realized, strikingly, even without the action of external magnetic fields. This is specifically demonstrated in a rectangular array of F-shaped meta-atoms patterned on a silicon dioxide (SiO$_2$). Notwithstanding this particular example, they provide with a general and effective pathway to implement their symmetry-based approach to engineer metasurfaces by breaking both mirror and inversion symmetries simultaneously so as to impart and control the spin (polarization-like) coherence in incoherently generated thermal radiation. Other relevant realizations are based on the possibility to actively switch the linear polarization of thermal fields. This has been numerically investigated and experimentally demonstrated in a MIM plasmonic structure by introducing a phase-changing material (GST) in the insulator layer, which allows for a rotation of the linear polarization enabled by the switching of the emissivity yielded by the transition between the phases amorphous and crystalline of the material~\cite{Qu2018A}. Another outstanding proposal consists in leveraging the capabilities of optical nanowire antennas to produce resonant excitation of highly polarized far-field thermal emission~\cite{Schuller2009}. Other than linear polarization, a resonant silicon-based chiral metasurface with broken mirror-inversion-symmetry has experimentally proven to be highly efficient for the generation of circularly polarized thermal radiation~\cite{Wu2014}. Likewise, there are other chiral structures, for example based on a layer-by-layer photonic crystal, where the circularly polarized thermal emission arises as a result of the polarization-dependent response within the photonic bandgap~\cite{Lee2007}. Such a realization has also been further optimized to emit narrowband circularly polarized thermal radiation~\cite{Dyakov2018}. Finally, there is a much more recent proposal to produce circularly polarized thermal emission from a compact dimer of subwavelength, anisotropic antennas, provided that they are out of thermal equilibrium, i.e., at different temperatures~\cite{Khandekar2019}.

\subsection{Radiative regimes of thermal emission}

Just like in conventional nanophotonic systems~\cite{Novotny}, according to the length from the radiation source, as well as the size of the emitter, in thermal emission engineering one can distinguish between two well differentiated radiative regimes with very distinct behaviors~[\hyperref[Fig.05]{Fig.~5(a)}]: the far and the near-field regimes. Regardless whether one is dealing with nanophotonics or thermal emission engineering, the former regime refers to propagating modes displaying an oscillatory behavior, whereas the latter means for sharply confined (either surface or guided) evanescent modes exhibiting an exponentially decaying behavior. Particularly in the realm of thermal emission, it has been demonstrated that both regimes exhibit coherence~properties that greatly differs from each other~\cite{Carminati1999,Shchegrov2000,Mulet2010,Edalatpour2016}. This directly translates into remarkable differences in the emission spectra, the directivity, and the polarization features of thermal fields.

As previously anticipated, such distinctions are essentially due to the existence of evanescent modes, which are dominant in the near-field and negligible in the far-field. This becomes neatly evident within~the~formalism of the {\em angular spectrum representation}~\cite{Novotny,Mandel}, often referred to as the {\em generalized plane-wave expansion}~\cite{Devaney1974}, a classical theoretical technique that enables a modal representation of any electromagnetic field in homogeneous media in terms of elementary plane waves, which can be either propagating or evanescent~\cite{Novotny,Mandel}. This treatment has proven to be specially well suited for analytically describing fields (or their propagators, namely, the corresponding dyadic Green's functions~\cite{Sipe1987,Francoeur2009,Narayanaswamy2014}) in material structures with planar geometries, such as slabs, interfaces, or layered media, wherein, due to the translational symmetry, the only relevant dimension is that pointing along the propagation direction. Hence, and without any loss of generality, assuming an electromagnetic (either optical or thermal) field propagating along the $z$-axis, and a wavevector defined as ${\bf k}=(k_x,k_y,k_z)$, so that $|{\bf k}|=k=n\omega/c$, with $n=\sqrt{\varepsilon\mu}$ being the refractive index, it can be demonstrated that, in the partial Fourier (or momentum) space~\cite{Goodman}, the field evolution along the $z$-axis can be simply expressed as:
\begin{equation}
\tilde{\boldsymbol{\Psi}}(k_x,k_y;z)=\tilde{\boldsymbol{\Psi}}(k_x,k_y;0)e^{\pm i k_z z},
\label{Eq.06}
\end{equation}
so that:
\begin{equation}
\boldsymbol{\Psi}(x,y,z)=\iint{dk_xdk_y \tilde{\boldsymbol{\Psi}}(k_x,k_y;0)e^{i\left(k_xx+k_yy\pm k_zz\right)}},
\label{Eq.07}
\end{equation}
where the sign $\pm$ indicates the sense of propagation, and:
\begin{equation}
k_z=\left\{\begin{matrix}
\sqrt{k^2-k_R^2}&\quad&\text{if}&\quad&k_x^2+k_y^2=k_R^2\leq k^2;\\
i\sqrt{k_R^2-k^2}&\quad&\text{if}&\quad&k_x^2+k_y^2=k_R^2>k^2.
\end{matrix}\right.
\label{Eq.08}
\end{equation}
This angular spectrum representation is applicable to both the electric and the magnetic fields, from which it can be straightforwardly noticed that the character of the wavevector $k_z$, which can be either real or imaginary, determines the behavior of the modes, which can be either oscillatory (propagating), or exponentially decaying (evanescent). This characterization becomes particularly simple in the case of lossless media, i.e., those for which $n$ is real and positive, where it is possible to establish the usual correspondence of $k_R\leq k$ and $k_R>k$, respectively, with propagating or evanescent modes. In this sense, this relatively simple mathematical description based on the relative position of the modes with respect to the lightcone $k=nk_0$, provides with a sharp physical characterization to precisely determine the radiative regime of thermal emission.

\subsubsection{Far-field thermal emission}

\begin{figure*}[t!]
	\centering
	\includegraphics[width=1\linewidth]{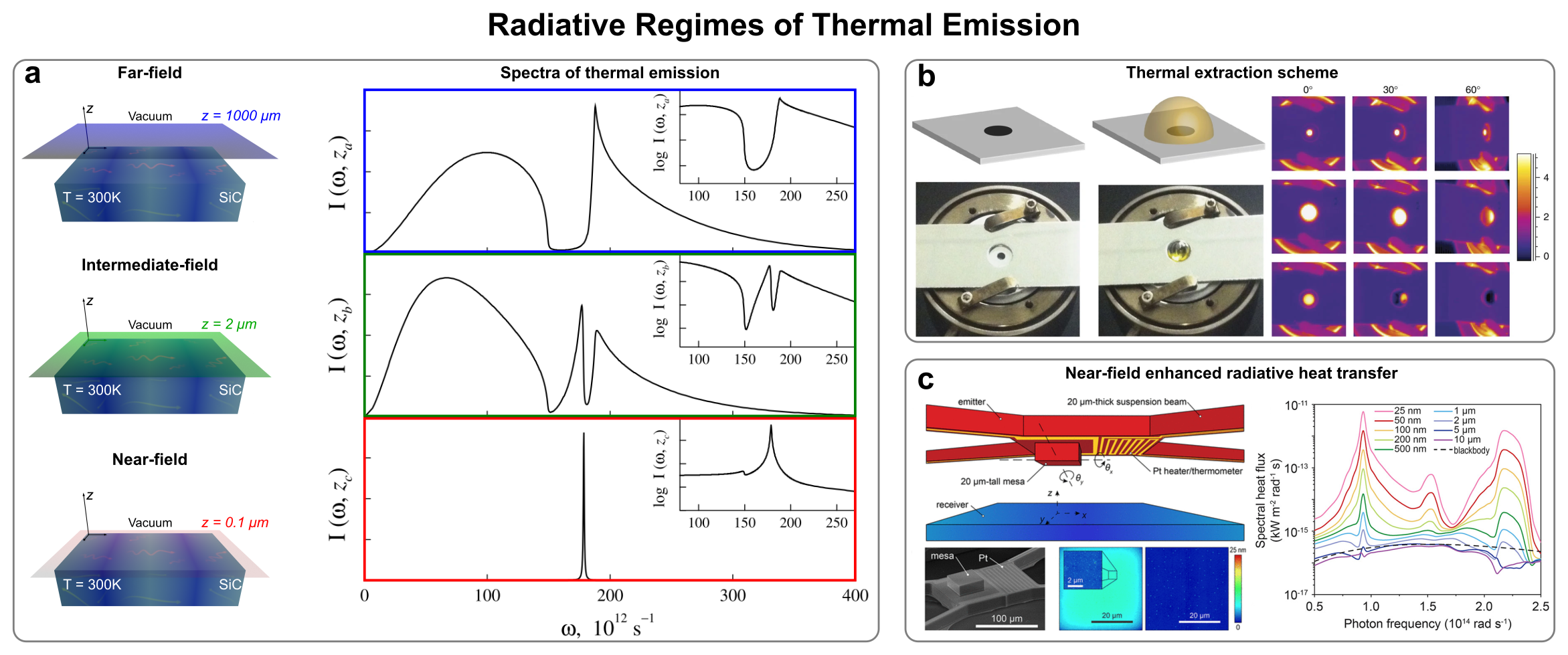}
	\caption{\textbf{Radiative regimes of thermal emission.} (\textbf{a})~Spectra of thermal emission evolves from far to near-field regimes displaying remarkable differences~\cite{Shchegrov2000}. (\textbf{b})~Thermal extraction enables a mechanism for tuning and/or enhancing the far-field thermal emission spectra relying on DOS~engineering~\cite{Yu2013}. (\textbf{c})~Experiments in nanometer-sized gaps have demonstrated that near-field radiative heat transfer can be extremely large, exceeding the blackbody limit by several orders of magnitude~\cite{Fiorino2018A}.}
	\label{Fig.05}
\end{figure*}

Far-field thermal emission is at the same time ruled and constrained by both Planck's and Kirchhoff's radiation laws. Upon this ground, aided by the aforementioned theoretical developments and practical implementations carried out from nanophotonic engineering approaches~\cite{Fan2017Review,Li2018Review,Baranov2019Review}, most of the efforts, aimed at tuning and weighting the thermal emission spectra, have been devoted to the search of mechanisms for, either widely or selectively, suppressing~\cite{Cornelius1999,Yeng2012,Arpin2013,Streyer2013,Narayanaswamy2004,Liu2011,Sakr2017,Sugimoto1994,Lin2003A,Han2007,Kats2012,Rinnerbauer2014,Kan2017} and/or enhancing~\cite{Celanovic2005,Yang2017,Mason2011,Ikeda2008,Asano2016,Zoysa2012,Diem2009,Askenazi2017} the material emissivity~\cite{Lin2000,Ilic2016,Lin2003B,Leroy2017,Li2018B}.

Yet, there is a specially insightful approach, ultimately based on DOS engineering, for enhancing the far-field thermal emission spectra by means of the modification of the refractive index of the surrounding medium: the {\em thermal extraction scheme}~\cite{Yu2013,Simovski2015}~[\hyperref[Fig.05]{Fig.~5(b)}]. Indeed, it should be noted that Planck's radiation law, as given in Eq.~\eqref{Eq.04}, refers to a blackbody emitter placed in a vacuum environment, i.e., in a surrounding medium with $n=1$. However, if we assume that the emitter is embedded in a transparent dielectric medium with a different refractive index, the speed of light in such a medium is no longer $c$, but $v=c/n$, and accordingly, the expression for the spectral energy density turns slightly modified by a factor~$n^2$. In this manner, high refractive index media enable a direct mechanism to broadly enhance the far-field thermal emission spectra. This approach has been demonstrated both experimentally, by means of an emitter adequately placed within a transparent semispherical dome made of ZnSe, with a refractive index of $n=2.4$~\cite{Yu2013}, and theoretically, by considering the usage of hyperbolic metamaterials~\cite{Simovski2015}. Here it must be emphasized that, inasmuch as the refractive index could be arbitrarily large, there is no theoretical limit on the maximum thermal emission enhancement achievable. Despite that, this does not mean at all that this thermal extraction scheme allows for surpassing the blackbody emission spectrum dictated by Planck's radiation law, since in such a case, it should be accounted for the entire system, including both the emitter and the surrounding medium as well.

\subsubsection{Near-field thermal emission}

Besides affording a higher performance of the far-field thermal emission, the nanophotonic approaches have also fostered the investigation of near-field thermal radiation, which has in turn boosted the development of a wealth of novel predictions, striking thermal effects, and innovative applications~\cite{Cuevas2018Review,Li2021Review,Pascale2023A,Chapuis2023}. Nonetheless, it is worth pointing out that, although most of the advances in this regime have certainly been carried out over the last few years, precisely due to the attainment of technical capabilities for the realization of complex photonic nanostructures, the analysis of thermal emission in the near-field regime (also referred to as {\em near-field radiative heat transfer}) is a long-standing issue whose first rigorous theoretical model is generally attributed to the seminal paper put forward by Polder and Van Hove in 1971~\cite{Polder1971}, and subsequent works by Pendry~\cite{Pendry1999}, and Volokitin~\cite{Volokitin2001}.

The central idea of near-field radiative heat transfer is that, in systems at a given finite temperature $T$, where the size and/or the separation distances are of the order of, or smaller than, the thermal wavelength (typically around $10~\mu{\rm m}$ at room temperature), the amount of thermal radiation emanating from the hot body can greatly exceed, even by several orders of magnitude, that predicted by Planck's radiation law~for~a~macroscopic blackbody in the far-field regime~\cite{Cornelius1999,Hu2008,Narayanaswamy2009,Bernardi2016}~[\hyperref[Fig.05]{Fig.~5(c)}]. Such an enhancement can be eventually attributed to the occurrence of interference effects due to multiple wave reflections in the gap between nearby objects~\cite{Basu2009A,Mulet2010}. But more prevalently, the enhancement is actually due to the presence of evanescent electromagnetic modes at the surface of materials as a consequence of their dispersive and absorptive features~\cite{FernandezHurtado2017,Iizuka2018}. In either case, this causes a significant increase of the LDOS, exhibiting a strong dependence with the geometrical features of the emitter, so that it may strongly exceed the DOS associated to the available thermal photons in the far-field regime~\cite{Babuty2013}. In general terms, such a spatial dependence can be encapsulated within the Green's function formalism~\cite{Joulain2003,Joulain2005}:
\begin{equation}
{\rm LDOS}=\frac{\omega}{\pi c^2}\text{Im}{\left\{\text{Tr}{\left[{\bf G}^{\rm E}({\bf r},{\bf r}',\omega)+{\bf G}^{\rm M}({\bf r},{\bf r}',\omega)\right]}\right\}},
\label{Eq.09}
\end{equation}
where the superscripts ${\rm E}$ and ${\rm M}$ stand, respectively, for the electric and magnetic contributions to the dyadic Green's function, which, ultimately, characterize the optical properties, and hence the response, of the medium. Still, for planar emitters there is a particular approximation, commonly known as the {\em quasistatic approximation}, whereby the Fresnel coefficients, and then the Green's functions, are simplified so that, in the asymptotic limit of modes with large wavevector~\cite{Rousseau2012}, strictly the limit $k\to\infty$, the LDOS near the material surface reduces to~\cite{Carminati1999,Shchegrov2000,Jones2013}:
\begin{equation}
{\rm LDOS}(d,\omega)=\frac{1}{16\pi^2\omega d^3}\frac{\text{Im}{[\varepsilon(\omega)]}}{|1+\varepsilon(\omega)|^2}.
\label{Eq.10}
\end{equation}
This asymptotic form of the LDOS highlights three pivotal insights of near-field thermal emission. First is the strong dependence with the distance $d$, which has been experimentally verified in several platforms with different architectures, reaching, and even surpassing, nanometre-scale distances~\cite{Rousseau2009,Nefedov2011,Gelais2014,Gelais2016,Kim2015,Cui2017,Fiorino2018A}~[\hyperref[Fig.05]{Fig.~5(c)}]. Second is that in the limit of purely lossless (and hence, dispersionless) media, since $\text{Im}{[\varepsilon(\omega)]}\to0$, thermal emission drastically vanishes, thereby revealing the essential relationship between dissipation and thermal emission processes. Noteworthily, this statement is also applicable in the far-field regime, which is neatly evinced within the fluctuational approach. And third, is that the above expression clearly underscores the crucial role of the (surface) resonant modes as enhancers of thermal emission~\cite{Kivisaari2022,Wang2016}. Indeed, it is easy to see that, at the frequencies for which $\varepsilon(\omega)=-1$, i.e., those associated with the resonant excitation~of~surface~polaritonic modes (either SPhPs~or~SPPs)~\cite{Biehs2013,LeGall1997,Shen2009,Ilic2012,Dai2015,Svetovoy2012,Liu2015D,Francoeur2008B,Zwol2011A,Castillo2022}, the ${\rm LDOS}$ displays a sharp peak~\cite{Basu2009B}. By~analyzing the evolution of the thermal emission spectrum, and hence of the DOS, as a function of the distance from the emitter, all these features have been rigorously demonstrated theoretically~\cite{Shchegrov2000,Joulain2005}, thereby explaining comprehensively the remarkable differences between the far and the near-field thermal emission spectra. Notwithstanding, following a similar approach as that yielding the far-field thermal extraction, various mechanisms have been proposed to extract and couple the enhanced near-field contribution to free-space propagating radiation~\cite{Messina2016,Ding2016,Li2019}, enabling so a further leverage of the unique features of near-field thermal emission, by transferring them into the far-field regime.

Near-field thermal emission has thoroughly been investigated in a plethora~of systems and configurations,~both~theoretically~\cite{Biehs2010,Iizuka2015} and experimentally~\cite{Volokitin2011C,Watjen2016}. Much more extensively, in order to tackle on systems with more complex geometries, it has also been addressed by means of numerical methods and simulations~\cite{Domingues2005,McCauley2012}, and inverse design approaches~\cite{Jin2019}. At any rate, despite the extraordinary enhancements predicted and reported, it should be noted that, akin to the far-field regime, there are fundamental upper bounds that limit the optical response in the near-field, and hence the radiative heat transfer, independently on the geometrical and dispersive features of the material~\cite{Miller2015,Shim2019,Molesky2020}. Furthermore, it is worth pointing out that, regardless of the enhancement, as long as one is dealing with thermal emission in the near-field regime, Planck's radiation law [Eq.~\eqref{Eq.04}] does not provide an appropriate description since, by construction, is inherently and solely associated to the far-field regime, and hence cannot be applied~\cite{Narayanaswamy2009}.

Beyond the fundamental interest in theoretically understanding, modeling, and experimentally proving the properties and limits of near-field thermal emission, these breakthroughs on near-field radiative heat transfer are also fostering the overhauling and upgrading of practical applications~\cite{Song2015,Lucchesi2021B}. In particular, an illustrative example lies on the aforementioned TPV systems, where recent works are showing that the inclusion of near-field thermal radiation effects brings about renewed insights to substantially improve the~thermal-to-electrical energy conversion efficiency~\cite{Zhao2017,Fiorino2018B,Inoue2019,Bhatt2020,Lucchesi2021A,Song2022}. More insightfully, following the same parallelism between electronics and photonics that inspired the development of optical analogues to lumped circuit elements~\cite{Engheta2007}, recently there is a groundbreaking proposal for devising near-field thermal analogues to the corresponding building blocks in electronic circuits. This innovative idea has led to the realization of thermal diodes~\cite{Fiorino2018C}, thermal transistors~\cite{BenAbdallah2014}, and solid-state thermal memories~\cite{Xie2011}, thereby giving rise to an emerging field termed~as~{\em thermotronics}~\cite{BenAbdallah2017}.

\subsubsection{Planck's and Kirchhoff's laws: Boundaries and breaches}

Overall in science and particularly in physics, the search and establishment of limits (either upper or lower) constitutes a paramount goal. On the one side, they sharply delimit and constraint the extension area of a research field. But, at the same time, they are often related with the existence of fundamental constants. In regards to the field of thermal radiation, this is clearly illustrated, e.g., by means of the Planck, the Boltzmann, the Stefan, and even the vacuum speed of light constants. Likewise, such limits may also be related with the bounded character in the value or even the trend of certain functions, e.g., the absolute zero of temperature, the emissivity (or the absortivity), restrained between~$0$~and~$1$, or the entropy increasement. Notwithstanding the foregoing, the establishment of limits is only a partial goal, since after that, the immediate question is whether is it possible to overcome them. In this sense, the eagerness for exploring borderlines of science, to some extent boosted by the continuous advance of the technical capabilities, is progressively fueling the upsurge in the search of mechanisms enabling to break down, and hence expand, these constraints. Sometimes, bounds are absolute, as in the case of the zero-point of temperature or the speed of light in vacuum, which are underpinned by essential characteristics that we assume as true, as the finite character of nature, or the principle of causality. However, more often, they are tied to artificial and ideal assumptions that we made for convenience and simplicity. This is precisely the case that justifies the so far established upper bound for Planck's and Kirchhoff's radiation laws, each of them being respectively associated to the thermal equilibrium condition and the reciprocity of the systems. Furthermore, alongside these assumptions, it is also crucial to seamlessly define the baseline conditions of the subject system, which for both, Planck's and Kirchhoff's laws, concern to the far-field regime of the radiation emitted by a macroscopic body~\cite{Hu2008,Narayanaswamy2009,Bernardi2016,Xiao2022}.

Upon this ground, significant efforts have been made to explore systems that stretch out such physical limits. Specifically, in order to enhance the far-field thermal emission, mechanisms such as thermal extraction~\cite{Yu2013,Shi2015,Messina2016,Ding2016,Li2019}, have been proposed. Likewise, it has been demonstrated that near-field thermal emission spectra can largely overcome the Planck's blackbody radiation law at the nanoscale~\cite{Hu2008,Narayanaswamy2009,Bernardi2016}. This latter possibility has brought about the introduction of the~term {\em super-Planckian} thermal emission, which has been extended to both~near and far-field regimes~\cite{Simovski2015,Biehs2013,Guo2012,Nefedov2014,Yang2018,FernandezHurtado2018A,FernandezHurtado2018B,Biehs2016,Maslovski2016,Thompson2018,Cuevas2019}. Inasmuch as the blackbody emission spectrum represents an upper limit, this feature would constitute a major breakthrough in the field of thermal emission engineering. However, very often, in both the near and far-field regimes, the use of such a term can hardly be adequately justified. On the one hand, it should be noted that Planck's law does not applies in the near-field regime, so the term super-Planckian~turns out to be directly meaningless~\cite{Biehs2013,Yang2018,Cuevas2019}. Yet,~even~in the far-field regime, the occurrence of super-Planckian emission mistakenly relies~on~the~use~of subwavelength emitters~\cite{Guo2012,Simovski2015,Biehs2016,Maslovski2016,FernandezHurtado2018A,FernandezHurtado2018B,Thompson2018,Nefedov2014}, which circumvents the applicability domain of Planck's law, as far as the size of the emitters is concerned~\cite{Xiao2022}. Indeed, due to the resonant behavior of subwavelength emitters, the absorption cross-section can largely exceed the geometrical cross-section of emitters~\cite{Ruan2010,Bohren1983}, resulting in an enhancement of the radiative heat transfer~that misleadingly appears to be super-Planckian~\cite{Xiao2022,Molesky2019}. Nonetheless, it is possible to overcome Planck's radiation law just by disregarding each of the underlying constraints, namely, the near-field regime, or the conditions of thermal equilibrium~\cite{Zou2021}. In particular, a typical approach to deal with non-equilibrium systems relies on the use of nonlinear media~\cite{Khandekar2015A,Khandekar2015B,Soo2016,Boyd,Khandekar2020}. Such is the case, for example, of a semiconductor externally biased either electrically or optically, which produces a redistribution of the energy of electrons and holes in different quasi-Fermi levels described by $qV_{\rm e}$ and $qV_{\rm h}$, where $q$ and $\Delta V=V_{\rm e}-V_{\rm h}$ stand, respectively, for the electron charge and the potential difference. This can be modeled by introducing a non-zero chemical potential, $\mu_{\rm F}=q\Delta V$, so that the spectral energy density of non-equilibrium thermal radiation is given~by~\cite{Wurfel1982,Ries1991,Herrmann2005}:
\begin{equation}
I_{\rm NE}(\omega,T,\mu_{\rm F})=\frac{\omega^2}{\pi^2c^3}\hbar\omega\left[\frac{1}{e^{(\hbar\omega\pm \mu_{\rm F})/k_BT}-1}\right].
\label{Eq.11}
\end{equation}
This expression, valid provided that $\hbar\omega\pm \mu_{\rm F}>k_BT$, slightly deviates from Planck's law on account of $\hbar\omega\pm \mu_{\rm F}$ [compare with Eq.~\eqref{Eq.04}], and provides with an extra degree of freedom for controlling the frequency distribution of thermal radiation. Thus, at a given temperature, photons with positive ($+\mu_{\rm F}$) or negative ($-\mu_{\rm F}$) chemical potential yield a higher or lower overall spectral energy density at every frequency. Besides being the basis of the aforementioned TPX technology~\cite{Harder2003B,Xue2015}, the higher control of radiative~heat transfer afforded by the chemical potential has enabled the theoretical proposal of novel thermal functionalities~\cite{Oksanen2015,Zhao2020}, such~as~near-field high-performance solid-state cooling~\cite{Sadi2020,Chen2015A,Chen2017}, and negative luminescent refrigeration~\cite{Chen2016A}, which has recently been experimentally demonstrated~\cite{Zhu2019}.

On the other side, the emissivity-absorptivity equivalence, established by the Kirchhoff's radiation law, relies on the reciprocity of systems~\cite{Greffet1998,Snyder1998}. In~turns,~akin~to the Planck's law, such a fundamental statement strongly roots on the assumption that the system should be in the far-field regime, including both the macroscopic size of the emitter and the observation distance. Noticeably, such a condition is precisely what underpins the bounded character of the emissivity, and, accordingly, that of the absorptivity. Indeed, just like the spectral emissivity of a material can be defined as the ratio between the spectral energy density of such an object and that of the blackbody [see Eq.~\eqref{Eq.01}], alternatively, the absorptivity can be defined as the absorption efficiency, i.e., as the ratio between the absorption and the geometrical cross-sections. Hence, since subwavelength objects often have absorption cross-sections much larger than geometrical cross-section, the absorptivity might be larger than $1$, and could even be completely unbounded, since an arbitrarily large absorption cross-section can be engineered~\cite{Ruan2010,Bohren1983}. Therefore, the use of subwavelength systems also circumvent the applicability domain of Kirchhoff's radiation law. Notwithstanding the foregoing, it is still possible to greatly violate the detailed balance~\cite{Zhu2014A}, and hence, overcome the Kirchhoff's radiation law in macroscopic emitters in the far-field regime by means of nonreciprocal materials~\cite{Hadad2016,Guo2020B,FernandezAlcazar2021}. Such a breakdown of reciprocity has been theoretically investigated in various systems, including semitransparent structures~\cite{Park2021}, magneto-optical materials~\cite{Zhao2019A}, spatio-temporally modulated media~\cite{Ghanekar2022,Torrent2018}, magnetic Weyl semimetals~\cite{Wu2021}, or gyrotropic materials~\cite{Fernandes2023}. Recently, this violation of Kirchhoff's law has also been experimentally observed in a system based on a guided-mode resonance coupled to a magneto-optic material~\cite{Shayegan2023}. Finally, it is worth highlighting two relatively recent works that, roughly speaking, generalize the treatment and extend the scope of validity of Kirchhoff's radiation law to both nonreciprocal~\cite{Miller2017}, and non-equilibrium systems~\cite{Greffet2018}.

Hence, upper bounds of both Planck's and Kirchhoff's laws are subjected to the constraints of the far-field regime, and the conditions of thermal equilibrium and reciprocity. In this sense, it is possible to overcome such fundamental laws just by disregarding each of those constraints, namely, undertaking the near-field regime, or breaking down the conditions of thermal equilibrium~\cite{Zou2021} or the reciprocity~\cite{FernandezAlcazar2021}. Yet, it is worth noticing that the most typical approaches to break down non-equilibrium, mainly based either on the use of nonlinear materials~\cite{Khandekar2015A,Khandekar2015B,Soo2016,Boyd,Khandekar2020}, or also dynamical (time-dependent) systems~\cite{Hadad2016,Coppens2017,Inoue2014,Ito2017,VazquezLozano2023A}, can also be used to break down the reciprocity~\cite{Sounas2017,Greffet2018}, which reveals such a close relationship between the notions of equilibrium and reciprocity.

\subsection{Theoretical frameworks for thermal emission}

As discussed in previous sections, Planck's and Kirchhoff's laws are generally regarded as the theoretical cornerstones of thermal radiation~\cite{Planck1901,Kirchhoff1860}. Nevertheless, it is crucial to realize that their applicability is inherently constrained to ideal systems fulfilling two quite sharp conditions. Firstly, the emitter should be amenable to be characterized as a blackbody, namely, as a non-reflective and totally absorptive macroscopic object in thermal equilibrium. Secondly, the whole system (including both the geometrical configuration and the material) must satisfy the principle of reciprocity~\cite{Onsager1931,Snyder1998,Greffet1998}. While such considerations are reasonable for addressing thermal emission in the far-field regime, i.e., that radiated by macroscopic emitters at large distances, they strongly fail in the near-field regime. This is essentially due to the need for taking into account the contribution of evanescent waves, as well as the possible occurrence of resonant responses of subwavelength emitters. In this regard, so as to provide a comprehensive and rigorous theoretical description of thermal emission, including both the far and the near field regimes, there are two particular frameworks which have proven to be very useful: {\em fluctuational electrodynamics}~\cite{Rytov} and {\em macroscopic quantum electrodynamics}~\cite{Vogel,Scheel2008}.

\subsubsection{Fluctuational electrodynamics}

Fluctuational electrodynamics~\cite{Rytov}, often referred to as stochastic electrodynamics~\cite{Marshall1963,Boyer1975,delaPena}, provides with a formidable theoretical framework to deal with classical electrodynamics under a statistical approach, thereby allowing us to tackle on electromagnetic systems, ultimately underpinned by Maxwell's equations, wherein fields are generated by randomly distributed and fleetingly moving sources: the fluctuating electromagnetic currents. The onset of such a classical formalism is generally attributed to the works carried out by Nyquist~\cite{Nyquist1928}, and by Callen and Welton~\cite{Callen1951}. Specifically, aimed at modeling the thermal noise in electrical circuits, in 1928, Nyquist derived an expression for the power spectral density of thermal voltage fluctuations in a resistor. Years later, in 1951, by recognizing that Nyquist's ideas could be applied more broadly to various physical systems, not just electrical circuits, Callen and Welton generalized the concept of thermal noise and connected it to linear response theory. Yet, and without a doubt, the main achievement of such a paper was the establishment of a connection between the equilibrium fluctuations (thermal noise) in a physical system and its linear response to small perturbations away from equilibrium. This relationship is now known as the {\em fluctuation-dissipation theorem}~(FDT)~\cite{Kubo1966}.

Since the FDT is a fundamental result of general scope, it has been formulated in many different manners. In the particular context of thermal emission, and referred to the current density correlations, it generally reads as:
\begin{equation}
\!\!\braket{{\bf j}^*({\bf r},\omega)\!\cdot\!{\bf j}({\bf r'},\omega')}_{\rm th}\!=\!4\pi\varepsilon_0\varepsilon''({\bf r},\omega)\hbar\omega^2\Theta(\omega,T)\delta{[\Delta{\bf r}]}\delta{[\Delta\omega]},
\label{Eq.12}
\end{equation}
where the brackets $\braket{\cdots}_{\rm th}$ denote a thermal ensemble~average, $\Theta(\omega,T)=\left[e^{\hbar\omega/(k_BT)}-1\right]^{-1}$, $\Delta{\bf r}={\bf r}-{\bf r}'$, $\Delta\omega=\omega-\omega'$, and $\varepsilon({\bf r},\omega)=\varepsilon'({\bf r},\omega)+i\varepsilon''({\bf r},\omega)$ stands for the dispersive and lossy permittivity of the material's emitter, i.e., the linear response function. Despite the diverse formulations, FDT generally provides a quantitative assessment of the correlations inherent to the fluctuating physical attributes of an equilibrium system, at the same time that establishes a close link between these correlations and the parameter that encapsulates the system's dissipative, or irreversible, and hence out-of-equilibrium, features, which are typically encompassed within the system's linear response function. Importantly, the FDT, as given in Eq.~\eqref{Eq.12}, seamlessly reveals the stochastic nature of thermal radiation. Indeed, this characteristic feature is directly reflected into the mathematics through the involvement of the Dirac delta functions, which ultimately underscore the uncorrelated character of thermally fluctuating currents at different positions and frequencies. Thus, the spectral energy density, yielding the spectrum of thermal radiation, can be reconstructed by adding the individual contributions of the fluctuating currents at each point of space, for each frequency, which is explicitly given by:
\begin{equation}
I_{\rm FDT}({\bf r},\omega,T)=\braket{{\bf E}^*({\bf r},\omega)\cdot{\bf E}({\bf r},\omega)}_{\rm th}.
\label{Eq.13}
\end{equation}
Then, noticing the connection between electromagnetic fields and currents:
\begin{equation}
{\bf E}({\bf r},\omega)=i\omega\mu_0\int_{\mathcal{V}}{d^3{\bf r}'{\bf G}^{\rm E}({\bf r},{\bf r}',\omega){\bf j}({\bf r',\omega})},
\label{Eq.14}
\end{equation}
where ${\bf G}^{\rm E}({\bf r},{\bf r}',\omega)$ is the dyadic Green's function of the body, it can be proved that:
\begin{equation}
I_{\rm FDT}=4\pi\mu_0\hbar\omega\Theta(\omega,T)\text{Im}{\left\{\text{Tr}{\left[{\bf G}^{\rm E}({\bf r},{\bf r},\omega)\right]}\right\}},
\label{Eq.15}
\end{equation}
where it has been used the completeness relation of the dyadic Green's function~\cite{Vogel,Novotny,Scheel2008}:
\begin{equation}
\text{Im}{\left[{\bf G}({\bf r},{\bf r}',\omega)\right]}\!=\!\frac{\omega^2}{c^2}\!\!\int_{\mathcal{V}}{\!\!d^3\boldsymbol{\rho}\varepsilon''(\boldsymbol{\rho},\omega){\bf G}({\bf r},\boldsymbol{\rho},\omega){\bf G}^*({\bf r}',\boldsymbol{\rho},\omega)},
\label{Eq.16}
\end{equation}
which can be derived from the {\em Schwarz reflection principle}, ${\bf G}^*({\bf r},{\bf r}',\omega)={\bf G}({\bf r}',{\bf r},-\omega^*)$, the {\em Lorentz reciprocity}, ${\bf G}^{\rm T}({\bf r},{\bf r}',\omega)={\bf G}({\bf r}',{\bf r},\omega)$, requiring the condition that ${\bf G}^*({\bf r},{\bf r}',\omega)\to0$ at ${\bf r}\to\infty$ (namely, ensuring that there is no net energy transport), and making use of the definition of ${\bf G}({\bf r},{\bf r}',\omega)$:
\begin{equation}
\nabla\times\nabla\times{\bf G}({\bf r},{\bf r}',\omega)-k^2{\bf G}({\bf r},{\bf r}',\omega)=\mathbb{I}\delta{[{\bf r}-{\bf r}']}.
\label{Eq.17}
\end{equation}
It should be noted that the FDT implicitly assumes the thermal character of the currents, and then, that of the fields. The resulting spectrum of thermal emission essentially depends on the characteristics of the emitter, specifically on its absorptivity (enclosed within the imaginary part of the permittivity), the temperature, as well as the geometrical aspects, including both the shape and size, described by the volume of integration~$\mathcal{V}$, and the position of observation~\cite{Carminati1999,Shchegrov2000,Joulain2005}. In this sense, Eq.~\eqref{Eq.15} not only stretches out the Planck's law for the blackbody radiation [compare with Eq.~\eqref{Eq.04}], but also recovers it in the particular case of emission into free-space, which can be readily verified just by properly identifying all the factors preceding the average energy distribution with the ${\rm DOS}$.

Inasmuch as it ultimately relies on classical electrodynamics, the above expressions stand for a semiclassical treatment of the FDT, where it is worth highlighting that, the vacuum contribution, represented by adding $1/2$ to the photon distribution, that is, $\Theta(\omega,T)\to\Theta(\omega,T)+1/2$, has been omitted. As pointed out in Ref.~\cite{Joulain2005}, such a consideration on whether to include or not the vacuum contribution is often rather arbitrary, in the sense that, from a classical approach, it is only based on heuristic arguments. This ambiguity becomes especially critical in the quantum context, where the vacuum (or zero-point) fluctuations are responsible of striking phenomena~\cite{Bimonte2009A,Messina2011,Kruger2012,Fong2019,Casimir1948,Lamoreaux1997,Dalvit,Jaffe2005,Rodriguez2011B,Woods2016}, such as the dynamical Casimir effect~\cite{Dodonov2020,Gong2021}, or, more generally, any other vacuum amplification effect~\cite{Nation2012,Renger2021}, as well as other quantum phenomena such as quantum friction~\cite{Volokitin1999,Kardar1999,Volokitin2011A,Silveirinha2014,Pendry1997,Philbin2009,Volokitin2011B,Philbin2011,Pendry2010A,Leonhardt2010,Pendry2010B,Zurita2004,Volokitin2007,Intravaia2014,Intravaia2016A,Intravaia2016B,Intravaia2019,Reiche2020,Oelschlager2022,Manjavacas2010A,Manjavacas2010B,Zhao2012,Manjavacas2017,Maghrebi2012,Maghrebi2014}. Fluctuational electrodynamics has proven to be a very successful framework that has made possible groundbreaking advances in thermal emission engineering via nanophotonic approaches~\cite{Polimeridis2015,Liberal2018,Francoeur2008A,Rodriguez2012,Rodriguez2013,Narayanaswamy2014}, including both the theoretical formalism and experimental platforms. However, as a semiclassical theory, it does not allow for the simultaneous modeling of quantum and thermal fluctuations. This reason should be sufficient to justify the need for addressing the theory of electromagnetic fluctuations (and hence, that of thermal emission) from a purely quantum approach~\cite{Mandel,Glauber1963,Mandel1965}. In fact, as stated by Glauber in Ref.~\cite{Glauber1963}, ``{\em it would hardly seem that any justification is necessary for discussing the theory of light quanta in quantum theoretical terms}''.

\subsubsection{Macroscopic quantum electrodynamics}

Broadly speaking, macroscopic quantum electrodynamics (QED) is a comprehensive theoretical formalism that extends quantum optics in free space to include the effects of absorbing and dispersive media~\cite{Scheel2008,Vogel}. Instead of bare photons, within this framework, one actually deals with elementary excitations, namely, quasiparticles or electromagnetic field-matter coupled states~\cite{Rivera2020}, represented by a continuum of harmonic oscillators. These are called polaritonic modes, and are described by means of quantum operators of creation, $\hat{\bf f}({\bf r},\omega_f;t)$, and annihilation, $\hat{\bf f}^\dagger({\bf r},\omega_f;t)$, which, in the Heisenberg picture, must fulfill the equal-time commutation relations:
\begin{subequations}
\begin{align}
&[\hat{\bf f}({\bf r},\omega_f;t),\hat{\bf f}({\bf r}',\omega_f';t)]=0,
\label{Eq.18a}\\
&[\hat{\bf f}^\dagger({\bf r},\omega_f;t),\hat{\bf f}^\dagger({\bf r}',\omega_f';t)]=0,
\label{Eq.18b}\\
&[\hat{\bf f}({\bf r},\omega_f;t),\hat{\bf f}^\dagger({\bf r}',\omega_f';t)]=\mathbb{I}\delta{[\Delta{\bf r}]}\delta{[\Delta\omega_f]},
\label{Eq.18c}
\end{align}
\end{subequations}
where $\mathbb{I}$ stands for the identity operator. Within this formalism, the dynamical behavior of a quantum photonic system can be described by means of a Hamiltonian~\cite{Vogel,Scheel2008}, $\hat{\mathcal{H}}=\hat{\mathcal{H}}_{\rm 0}+\hat{\mathcal{H}}_{\rm int}$, where $\hat{\mathcal{H}}_{\rm 0}$ and $\hat{\mathcal{H}}_{\rm int}$, characterize, respectively, the polaritonic (light-matter coupled) environment, and the interaction yielded by a polarization field induced by an external electric field~\cite{Loudon,Boyd}:
\begin{subequations}
\begin{align}
&\hat{\mathcal{H}}_{\rm 0}=\int{d^3{\bf r}\int_{0}^{+\infty}{d\omega_f \hbar\omega_f\hat{\bf f}^\dagger({\bf r},\omega_f;t)\cdot\hat{\bf f}({\bf r},\omega_f;t)}},
\label{Eq.19a}\\
&\hat{\mathcal{H}}_{\rm int}=-\int{d^3{\bf r}\hat{\boldsymbol{\mathcal{P}}}({\bf r},t)\cdot\hat{\boldsymbol{\mathcal{E}}}({\bf r},t)}.
\label{Eq.19b}
\end{align}
\end{subequations}
Accordingly, the polarization field operator is generally expressed as $\hat{\boldsymbol{\mathcal{P}}}({\bf r},t)=\int_{0}^{t}{d\tau \Delta\chi({\bf r},t,\tau)\hat{\boldsymbol{\mathcal{E}}}({\bf r},\tau)}$~\cite{Novotny,Boyd}, where $\Delta\chi$ is the susceptibility, or electric response, function, that, in general, depends on both space and time. Furthermore, it is worth noticing the mathematical character of the electric and polarization vector fields, which are not functions, but quantum operators (denoted with a hat), thus bearing well known properties such as their way to be applied over quantum states, or their, in general, non-commutative character. In this regard it should be noted that the electric field operator generally reads as, $\hat{\boldsymbol{\mathcal{E}}}({\bf r},t)=\hat{\boldsymbol{\mathcal{E}}}^{(+)}({\bf r},t)+\hat{\boldsymbol{\mathcal{E}}}^{(-)}({\bf r},t)$, where:
\begin{equation}
\!\!\!\!\hat{\boldsymbol{\mathcal{E}}}^{(+)}({\bf r},t)=\int_{\mathcal{V}}d^3{\bf r}'\int_{0}^{+\infty}{\!\!\!\!d\omega_f {\bf G}^{\rm E}({\bf r},{\bf r}',\omega_f)\hat{\bf f}({\bf r}',\omega_f;t)},
\label{Eq.20}
\end{equation}
with:
\begin{equation}
\!\!\!\!{\bf G}^{\rm E}({\bf r},{\bf r}',\omega_f)\!=\!i\sqrt{\frac{\hbar}{\pi\varepsilon_0}}\left[\frac{\omega_f}{c}\right]^{\!2}\!\!\sqrt{\!\varepsilon''({\bf r}',\omega_f)}{\bf G}({\bf r},{\bf r}',\omega_f),\!\!
\label{Eq.21}
\end{equation}
being the response function, depending on the dyadic Green's function ${\bf G}({\bf r},{\bf r}',\omega_f)$, characterizing the medium, and noticing that $\hat{\boldsymbol{\mathcal{E}}}^{(-)}({\bf r},t)=[\hat{\boldsymbol{\mathcal{E}}}^{(+)}({\bf r},t)]^\dagger$.

Within this quantum framework, the thermal emission spectrum can be obtained from the electric field correlation~\cite{Mandel,Glauber1963,Mandel1965}:
\begin{equation}
I_{\rm MQED}=\braket{[\hat{\bf E}^{(+)}({\bf r},\omega)]^\dagger\cdot\hat{\bf E}^{(+)}({\bf r},\omega)}_{\rm th},
\label{Eq.22}
\end{equation}
where $\hat{\bf E}^{(+)}({\bf r},\omega)=\mathcal{L}_{\omega}[\hat{\boldsymbol{\mathcal{E}}}^{(+)}({\bf r},t)]$ is the Laplace's transform of the electric field operator. This expression resembles to great extent that given within the fluctuational approach~[compare~with~Eq.~\eqref{Eq.13}]. However, inasmuch as the positive- and the negative-frequency parts of the fields are respectively associated to the annihilation and creation polaritonic operators, it explicitly emphasizes the need of carefully regarding the ordering (either normally or anti-normally~\cite{Agarwal1975,Mandel1966}) to properly distinguishing between the photon absorption and emission processes~\cite{Vogel,Scheel2008}. It must be noted that thermal expectation values of electromagnetic fields are directly tied from those of the polaritonic operators. Indeed, by assuming electromagnetic fields in thermal equilibrium at temperature $T$ (i.e., thermal fields), it can be demonstrated that~\cite{Scheel2008}:
\begin{subequations}
\begin{align}
&\braket{\hat{\bf f}^\dagger({\bf r},\omega)\cdot\hat{\bf f}({\bf r}',\omega')}_{\rm th}=\Theta(\omega,T)\delta{[\Delta{\bf r}]}\delta{[\Delta\omega]},
\label{Eq.23a}\\
&\!\braket{\hat{\bf f}({\bf r},\omega)\cdot\hat{\bf f}^\dagger({\bf r}',\omega')}_{\rm th}=[\Theta(\omega,T)+1]\delta{[\Delta{\bf r}]}\delta{[\Delta\omega]},
\label{Eq.23b}
\end{align}
\end{subequations}
where $\braket{\cdots}_{\rm th}=\text{Tr}{[\cdots\hat\varrho_{\rm th}]}$, with $\varrho_{\rm th}$ being the thermal density operator~\cite{Vogel,Loudon}. The aforementioned relationships, along with the quantum version of the relation between electromagnetic fields and currents given in Eq.~\eqref{Eq.14}, afford a direct route for elucidating the quantum analogue of the FDT. Notably, by virtue of these established connections, the resulting FDT effectively removes the ambiguity surrounding the inclusion (or omission) of the quantum vacuum~contribution.

By considering different interaction Hamiltonians, $\hat{\mathcal{H}}_{\rm int}$, the formalism of macroscopic QED has found applications in several different contexts apart from thermal emission. In particular, it has facilitated the modeling and analysis of quantum emitters interacting with plasmonic systems~\cite{Delga2014}, as well as with resonant cavities and waveguides~\cite{Liberal2019}. Furthermore, macroscopic QED has also been crucial in elucidating dynamical vacuum amplification effects in time-varying optical media~\cite{Sloan2021}. Building upon this basis, it has been recently demonstrated that, a proper treatment of the interaction Hamiltonian paves the way for a comprehensive theoretical framework to bridge two fundamental and currently very active areas of research in the fields of nanophotonic engineering and physics, namely, time-varying media and thermal emission~\cite{VazquezLozano2023A}.

\section{Time-dependent thermal emission: dynamic tuning of thermal features and temporal metamaterials}

In the realm of thermal emission engineering, the control and manipulation thermal radiation, specifically the coherence properties (bandwidth, directivity, and polarization) both in the far and the near-field regimes, has hitherto been primarily based on passive methodologies. This entails the consideration of emitters, photonic platforms, and environments, characterized by fixed properties that remain static over time. A thorough examination of the existing literature underscores the prevalence of this passive approach both in photonic and thermal emission engineering. However, a notable shift toward active mechanisms, i.e., those involving the time as an extra degree~of~freedom~to~be exploited, is currently underway~\cite{Picardi2023Review,Inoue2014,Engheta2021,Galiffi2022,Yin2022,Yuan2022,Engheta2023}. Such an active approach enables dynamic control over thermal emission features, which besides granting access to fundamental~insights~tied to the breakdown~of~equilibrium~\cite{Xiao2022,Greffet2018,Dedkov2017,Dedkov2020,Ito2017},  or the reciprocity~\cite{Sounas2017,FernandezAlcazar2021,Torrent2018,Fernandes2023}, and~the~time reversibility~\cite{Tolman1948,Onsager1931,Miller1960,Ghanekar2022,Ries1983,Benenti2011,Callen1951,Callen1952}, it is also providing the system with some practical benefits, such~as a higher design~flexibility and reconfigurability~\cite{Coppens2017,Liu2017A,Phan2013,Vassant2013,Ihlefeld2015,Cao2018,Park2018,Wasserman2019,Wasserman2019}. This departure from conventional strategies constitutes a substantive leap in the field, offering new prospects for advancing our understanding of thermal radiation as well as improving the current capabilities of thermal emission engineering. Yet, it is worth noticing a subtle distinction between two different possibilities to deal with time-dependent thermal emission: an approach based on {\em dynamical tuning of thermal features}, mostly concerning the variation of the environments and the photonic platforms~[\hyperref[Fig.06]{Fig.~6(a)}], and other based on {\em temporal metamaterials} (often referred to as time-varying, or time-modulated, media), where the temporal dependence involves the own material properties of the emitters~[\hyperref[Fig.06]{Fig.~6(b)}].

\subsection{Thermal emission in dynamically tunable systems}

\begin{figure*}[t!]
	\centering
	\includegraphics[width=0.915\linewidth]{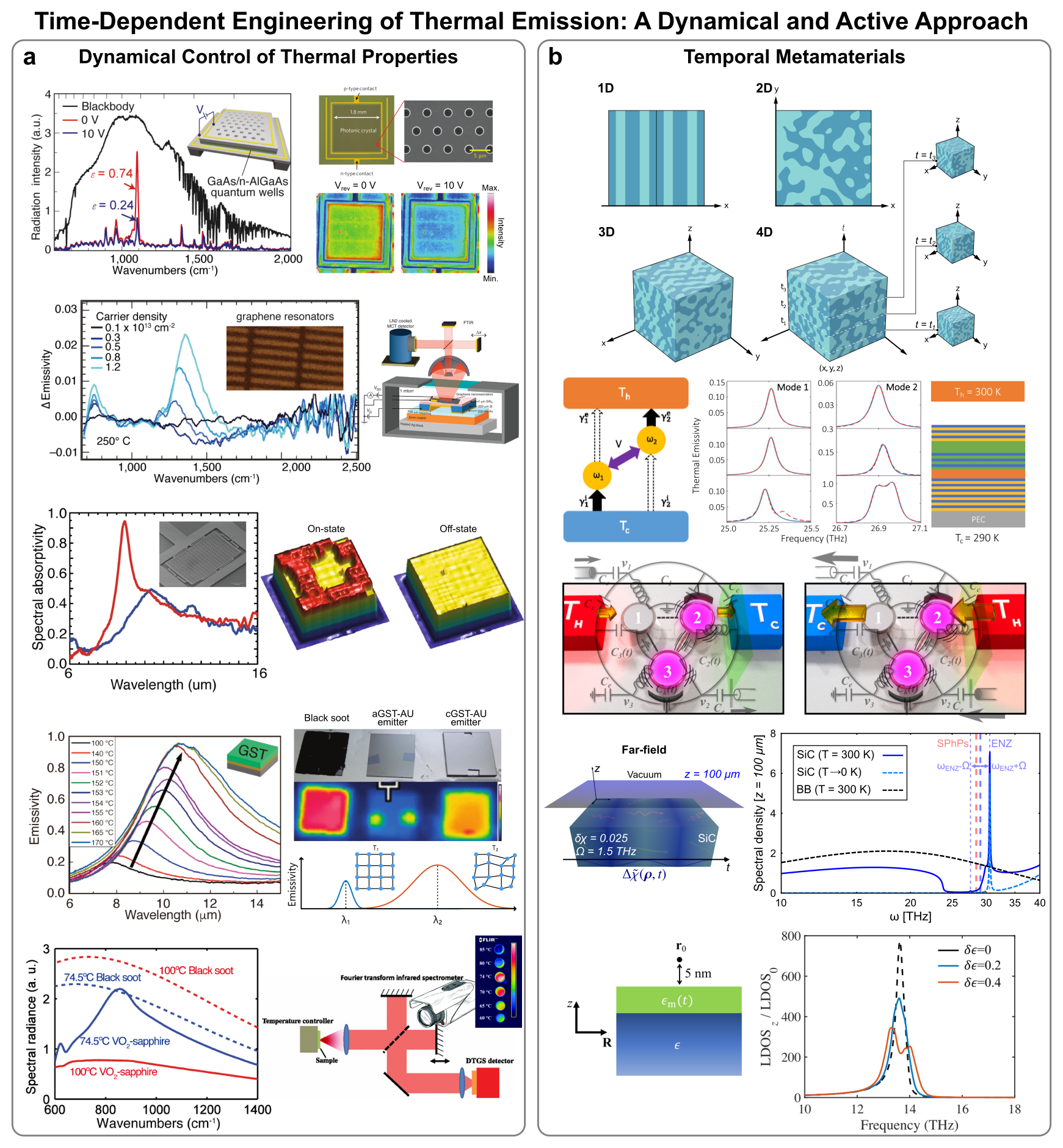}
	\caption{\textbf{Time-dependent thermal emission engineering.} (\textbf{a})~Palette of different tunable systems and configurations that have been theoretically and experimentally investigated to perform dynamic control of thermal emission~\cite{Inoue2014,Brar2015,Liu2017A,Du2017,Kats2013}. (\textbf{b})~By introducing the time as an additional and fundamentally different degree of freedom, temporal metamaterials (materials with a designed temporal modulation of their constitutive parameters) are revolutionizing the fields of optics and photonics engineering~\cite{Engheta2023}. Recent investigations are proving that this approach can also be used in the context of thermal emission engineering, showing innovative functionalities~\cite{FernandezAlcazar2021,Buddhiraju2020}, and extraordinary far and near-field thermal features~\cite{VazquezLozano2023A,Yu2023A}.}
	\label{Fig.06}
\end{figure*}

In order to further expand and improve the capabilities of thermal emitters to enable an active and real-time control, recent investigations are looking into the insightful opportunities of dynamically tuning thermal emission features~\cite{Picardi2023Review}. Akin to conventional passive approaches, the main objective is manipulating the emissivity (and/or the absorptivity) of the emitters and their surroundings. Besides a higher design flexibility, such an active approach provides with the possibility of thermal radiation to be real-time controlled, and hence, reconfigurable. As can be drawn from the aforementioned theoretical frameworks, the parameters that are susceptible to be (both passively and actively) tuned to modulate the thermal emission features essentially are the gradient of temperature from the emitter and the background medium, $\Delta T=T_{\rm em}-T_{\rm bg}$, and the dyadic Green's functions, entailing both the constitutive parameters, i.e., either the permittivity, $\varepsilon_{\rm em}(\omega)$, or the permeability, $\mu_{\rm em}(\omega)$, of the emitting body, and the refractive index of the surrounding medium, $n_{\rm bg}(\omega)$. Roughly speaking, the variation of the temperature is tied to a modification of the photon's frequency distribution, and the dyadic Green's function to engineering the frequency distribution of photon DOS. Accordingly, several dynamical tuning mechanisms have been theoretically and experimentally explored~[\hyperref[Fig.06]{Fig.~6(a)}], from electrostatic~gating~\cite{Vassant2013,Ihlefeld2015,Wasserman2019,Park2018,Xiao2019B,Burokur2010,Jun2013,Li2011,Wu2019A,Shrekenhamer2013,Brar2015,Huang2014}, electro-mechanical~stretching~\cite{Liu2017A,Moridani2017,Xu2018}, to thermo-optical~\cite{Qu2017,Kats2013,Kou2018,Nguyen2021}, and magneto-optical modulations~\cite{Moncada2015,Kollyukh2005,Wu2019B,Ekeroth2018}. Specially within this dynamical context, it is worth noticing that a crucial parameter to take into account is~the~emission and modulation speeds~\cite{Yu2017,Inoue2014,Wasserman2019,Xiao2019B,Zwol2011B,Sakat2018,Mori2014}, in terms of which, the best performance is generally reached for optical mechanisms~\cite{Lui2010,Xiao2019B}, followed by the electrical, mechanical, and thermal approaches. Likewise, regarding materials and platforms, quantum wells~\cite{Inoue2014,Vassant2013}, plasmonic metasurfaces~\cite{Park2018,Moridani2017}, graphene-based resonators~\cite{Brar2015}, and phase-change~materials~\cite{Cao2018,Kats2012}, such as GST~(Ge$_2$Sb$_2$Te$_5$)~\cite{Qu2017,Qu2018A,Du2017,Chen2013,Qu2018B} and vanadium dioxide (VO$_2$)~\cite{Sun2019,Zheng2022,Wang2014,Zwol2011A,Xie2011,Kats2013}, whose behavior depend on a structural (amorphous/crystalline) and/or electronic (dielectric/metallic) phase, which can be dynamically switched back and forth (e.g., via temperature variations or applying an external electric field), have proven to be promising candidates to actively manipulate thermal emission features both in the~far~\cite{Giteau2023B,Ito2017}, and, more recently, also in the near-field~regime~\cite{Thomas2019,Ge2019,Zwol2011A,Papadakis2019}.

In~\hyperref[Fig.06]{Figure~6(a)}, we show some illustrative examples of different tunable systems and configurations for dynamic modulation of thermal emission features. In particular, based on the electrostatic gating, one of the most representative realizations is that put forward by Inoue and colleagues~\cite{Inoue2014}, wherein, by means of a photonic crystal cavity coupled with multiple GaAs/n-AlGaAs quantum wells, they experimentally demonstrate fast dynamic control of thermal emissivity via modulation of the intersubband absorption in the quantum well, induced by the application of an external electric gate voltage which varies the charge carrier density. Upon a similar approach, this concept of gate-tunability has also been demonstrated in an array of graphene plasmonic resonators~\cite{Brar2015}. There are other proposals integrating metamaterial with microelectromechanical systems (MEMS) that harness the electric bias to modify the geometry of structures, and thus the emissivity~\cite{Liu2017A}. Finally we depict two particular examples enabled by phase-change materials allowing for a change in the emissivity tied to a phase switching. One is based on the structural transition between an amorphous and a crystalline phase of GST, which modifies the refractive index~\cite{Du2017}. Likewise, it has also been shown that the electronic transition between an insulator and metal phase of VO$_2$ enables a temperature-dependent mechanisms for actively engineering and tuning the emissivity~\cite{Kats2013}. Despite the above selected examples, presently, the actual challenge lies in the experimental validation of many other configurations theoretically proposed, as well as the development and implementation of feasible and practical applications, such as radiative~heat~management~\cite{Zhang2019,Tong2015}, thermal camouflaging~\cite{Qu2018B,Xiao2015,Phan2013}, self-adaptive radiative cooling~\cite{Ono2018,Xu2022}, and radiative thermal rectification~\cite{Yang2013,Ito2014,Chen2021}.

\subsection{Thermal emission from temporal metamaterials}

Regarding the active control and manipulation of thermal radiation in time-dependent systems, the latest breakthrough in thermal emission engineering has come with the recent advent of temporal metamaterials, also known as time-varying, or time-modulated, media. By providing an additional and fundamentally different degree of freedom~\cite{Engheta2021}, temporal metamaterials are being postulated as an enticing platform for actively and dynamically engineering optical properties and light-matter interactions~\cite{Galiffi2022,Yin2022,Yuan2022}, and are currently becoming in one of the more active areas of research in the fields of optics and nanophotonics~\cite{Engheta2023}.

This approach markedly differs from the above dynamical tuning mechanisms~\cite{Picardi2023Review}, both theoretically and practically. Indeed, roughly speaking, dynamical tuning methods involve the modulation of extrinsic properties, such as changes in temperature, phase transitions, or manipulations affecting to mechanical, electrical, or chemical properties. In contrast, temporal metamaterials, entail the temporal modulation of intrinsic constitutive properties, i.e., the permittivity and/or the permeability (commonly encompassed by the electromagnetic susceptibility). The underlying idea is somehow similar to the traditional conception of metamaterials~\cite{Heber2010,Soukoulis2011,Zheludev2010,Zheludev2015}, but, instead of (or in addition to) spatially modeling, the material properties are engineered by means of temporal modulations~\cite{Engheta2021,Galiffi2022,Yin2022,Yuan2022,Engheta2023}. In this manner, unlike conventional metamaterials, where the artificially designed material properties are persistent in time, in temporal metamaterials they are, by definition, inherently tied to the prevalence of a temporal modulation over the constitutive parameters characterizing the response of matter, which, in general, should be externally driven.

Empowered by the introduction of the dimension of time, temporal metamaterials has meant a qualitative leap in the field of nanophotonic engineering~[\hyperref[Fig.06]{Fig.~6(b)}], upgrading and enriching the variety of the achievable physical effects, phenomena, and applications~\cite{Caloz2020A,Caloz2020B,Taravati2020,Hayran2023}. Akin to previously sketched approaches and platforms, employed for controlling and enhancing the coherence properties of thermal radiation, the conceptualization of temporal metamaterials can also be exported to the field of thermal emission engineering, both in far and near-field regimes~\cite{Torrent2018,VazquezLozano2023A,Yu2023A,Yu2023B,VazquezLozano2023B}. However, if the topic of time-varying media is still at a very incipient stage in the field of nanophotonics engineering, even more so in the realm of thermal emission engineering, where references, and particularly in the experimental ground, are rather scarce. Notwithstanding, in \hyperref[Fig.06]{Figure~6(b)}, we showcase a selection of few recent works illustrating the theoretical potential of time-modulated media for controlling thermal radiation properties. Specifically, it can be seen a schematic depiction of the general setup enabled by the temporal modulation which can be used to yield a photon-based active cooling mechanism, namely, a thermal photonic refrigerator able to pump heat from a low-temperature to a high-temperature reservoir~\cite{Buddhiraju2020}. Following a similar approach, it has also been theoretically proposed a Floquet-based (time-varying) thermal diode leading to extreme nonreciprocal near-field thermal radiation~\cite{FernandezAlcazar2021}. Yet, it has not been until very recently, that a rigorous theoretical basis for studying thermal emission in time-modulated materials has been put forward~\cite{VazquezLozano2023A}. Such a formalism, developed under the framework of macroscopic QED, demonstrates that the temporal modulation gives access to new and extraordinary physical features of thermal emission~\cite{VazquezLozano2023B}; from the emergence of nonlocal correlations in space and frequency, to the occurrence of a sharp peak, at the material's ENZ frequency, in the far-field thermal emission spectrum, exceeding the blackbody. Interestingly, such a super-Planckian emission is  attributed to the dynamical amplification of quantum vacuum~\cite{Xiao2022} and is persistent at all regimes, which suggests an alternative ENZ-based thermal extraction scheme, simultaneously boosting near and far field thermal processes. Remarkably, this quantum formalism is entirely valid for addressing both far and near-field thermal emission. Still, a similar study has been performed from a fully classical approach based on fluctuational electrodynamics, showing that spatial coherence, tied to the directivity of thermal fields, may also be actively manipulated by means of time-modulated photonic structures~\cite{Yu2023A}. Furthermore, upon this same framework, it has also been theoretically demonstrated how the effects of time-modulation can result in the enhancement, suppression, or reversal of near-field radiative heat transfer between two~bodies~\cite{Yu2023B}.

According to the above theoretical studies, the new features brought about by time-varying media in the field of thermal emission engineering hold the promise of new avenues toward an enhanced and dynamical coherence control of thermal radiation. This may in turn led to enhance and upgrade into dynamically active conventional thermal applications and functionalities, including heat and energy management and harvesting~\cite{Bierman2016}, light sources~\cite{Ilic2016}, sensing~\cite{Lochbaum2017}, communications~\cite{Boriskina2017}, radiative cooling~\cite{Raman2014}, thermoregulation~\cite{Wei2021}, thermal camouflaging~\cite{Li2018A}, and imaging~\cite{Tittl2015}, among many others. Furthermore, the possibility of overcoming the blackbody spectrum and actively boosting thermal emission suggests the feasibility to perform radiative heat engines~\cite{Giteau2023A,Xiao2022}. Finally, it is also worth emphasizing that the introduction of the temporal degree of freedom offers an alternative method to modeling, or patterning, the material properties, circumventing the need of complex nanofabrication processes~\cite{Galiffi2020}, thereby removing all the costs, times, and technical limitations, along with the potential capability to achieve final outcomes with arbitrarily high levels of quality. Notwithstanding the foregoing, the current challenge of thermal emission in time-varying media lies in the experimental implementation, where recent progress in nanophotonics poses promising prospects~\cite{Galiffi2022}.

\section{Conclusions and Outlook}

The emission of thermal radiation stands out as one of the few singular processes that brings together both a fundamental and universal nature. This assertion can be readily understood from the following realizations: (1) it solely depends on the existence of a body at finite temperature, (2) it does not require a material medium for propagation, and (3) as dictated by the third principle of thermodynamics, the impossibility of a closed system by any finite physical procedure, no matter how idealized, to reach the absolute zero of temperature. At the same time, its consequences, both at fundamental and applied levels, widely span across many of the major scientific disciplines, such as chemistry, biology, and, of course in physics. Particularly in physics, it plays a central role over various of the main branches: thermodynamics, electrodynamics, and quantum theory. This is clearly illustrated from a historical standpoint by noticing the great amount of renowned and pioneering names that strongly contributed in laying down the foundations of thermal radiation, including both the theoretical comprehension and the experimental control. From the early works of Carnot, often dubbed as the ``father of thermodynamics'', scientists such as Kirchhoff, Lord Kelvin, Stefan, Boltzmann, Lord Rayleigh, Jeans, Wien, to Planck, and more indirectly, though equally essential, Herschel and Maxwell, among many others, they all have underpinned the field of thermodynamics, put forward the theoretical groundwork to deal with thermal radiation as a propagating electromagnetic wave, and, remarkably, paved the way toward the introduction of the quantum theory of light. Essentially, this conforms the pillars of what we nowadays know as thermal~emission~engineering.

Thermal emission engineering has undergone a substantial evolution, ultimately emerging as a subject of eminently interdisciplinary nature. As delineated throughout this review, currently there is a conspicuous and advantageous cross-disciplinary dynamic interrelation among communities engaged in the investigation of thermal radiation, whether from the standpoints of thermodynamics or quantum physics, and those dedicated to nanophotonic and material engineering. This convergence highlights the mutually beneficial interplay and synergies between traditionally distinct domains, underscoring the intricate integration of thermal emission engineering across diverse scientific and technological disciplines.

Upon this ground, latest advances carried out in the fields of optics and nanophotonics have fostered and also proven the crucial role of artificial nanostructures to control and enhance the coherence properties of thermal radiation, both in the far and the near-field regimes. At the same time, such developments have served as a guideline to upgrade the platforms and theoretical approaches in thermal emission. In particular, it has motivated the search of feasible mechanisms to overcome the Planck's and Kirchhoff's radiation laws, the investigation of time-dependent mechanisms to tune and dynamically modulate thermal emission features, as well as in the quantum realm, unifying the treatment to deal with thermal and quantum vacuum fluctuations. This is paving the way toward extraordinary thermal effects, and is already suggesting the exploration of novel technological applications. These strides are boosting the research in thermal emission engineering, thereby auguring a flourishing future of this field, again booming.

\section*{Author information}

\subsection*{Corresponding Author}
\vspace{-0.15cm}

\noindent\href{enrique.vazquez@unavarra.es}{enrique.vazquez@unavarra.es}\\
\href{inigo.liberal@unavarra.es}{inigo.liberal@unavarra.es}

\vspace{-0.35cm}
\subsection*{ORCID}
\vspace{-0.15cm}

\noindent J. Enrique V\'azquez-Lozano: \href{https://orcid.org/0000-0001-6423-1949}{0000-0001-6423-1949}\\
I\~nigo Liberal: \href{https://orcid.org/0000-0003-2620-6392}{0000-0003-2620-6392}

\vspace{-0.35cm}
\subsection*{Notes}
\vspace{-0.15cm}

The authors declare no competing financial interest.

\vspace{-0.35cm}
\section*{Acknowledgments}
\vspace{-0.15cm}

This work was supported by ERC Starting Grant No.~ERC-2020-STG-948504-NZINATECH. 
J.E.V.-L. acknowledges support from Juan de la Cierva--Formaci\'on fellowship~FJC2021-047776-I. I.L.~further acknowledges support from Ram\'on~y~Cajal fellowship~RYC2018-024123-I.

\end{document}